\documentclass[10pt]{article}
\usepackage[top=1in,bottom=1in,left=1in,right=1in]{geometry}
\usepackage{graphicx}
\usepackage{cite}
\usepackage{amsmath,amssymb,amsfonts}
\usepackage{caption}
\usepackage{physics}
\usepackage{authblk}

\title{Phase-Binarized Spintronic Oscillators for Combinatorial Optimization, and Comparison with Alternative Classical and Quantum Methods}

\author[1]{Neha Garg}
\author[2]{Sanyam Singhal}
\author[3]{Nakul Aggarwal}
\author[4]{Aniket Sadashiva}
\author[1]{Pranaba K. Muduli}
\author[4]{Debanjan Bhowmik}
\affil[1]{Department of Physics, Indian Institute of Technology Delhi}
\affil[2]{Department of Physics, Indian Institute of Technology Bombay}
\affil[3]{Department of Physics, University of Alberta, Canada}
\affil[4]{Department of Electrical Engineering, Indian Institute of Technology Bombay}

\date{}

\begin{document}
\maketitle




\begin{abstract}
Solving computationally intensive combinatorial optimization problems efficiently through emerging hardware by converting the problem to its equivalent Ising model and obtaining its ground state is known as Ising computing. Phase-binarized oscillators (PBO), modeled through the Kuramoto model, have been proposed for Ising computing, and various device technologies have been used thus far to experimentally implement such PBOs. In this paper, we show that an array of four dipole-coupled uniform-mode spin Hall nano oscillators (SHNOs) can be used to implement such PBOs and solve the NP-Hard combinatorial problem MaxCut on 4-node complete weighted graphs. We model the spintronic oscillators through two techniques: an approximate model for coupled magnetization dynamics of spin oscillators, and Landau Lifshitz Gilbert Slonckzweski (LLGS) equation-based more accurate magnetization dynamics modeling of such oscillators. Next, we compare the performance of these room-temperature-operating spin oscillators, as well as generalized PBOs, with two other alternative methods that solve the same MaxCut problem: a classical approximation algorithm, known as Goemans-Williamson's (GW) algorithm, and a Noisy Intermediate Scale Quantum (NISQ) algorithm, known as Quantum Approximation Optimization Algorithm (QAOA). For four types of graphs, with graph size up to twenty nodes, we show that approximation ratio (AR) and success probability (SP) obtained for generalized PBOs (Kuramoto model), as well as spin oscillators, are comparable to that for GW and much higher than that of QAOA for almost all graph instances. Moreover, unlike GW, the time to solution (TTS) for generalized PBOs and spin oscillators does not grow with graph size for the instances we have explored. This can be a major advantage for PBOs in general and spin oscillators specifically for solving these types of problems, along with the accuracy of solutions they deliver.  
\end{abstract}

%
%
%
%
%


\section{Introduction}

\subsection{Motivation}

Solving combinatorial optimization problems is useful for various practical applications: delivery scheduling, flight booking, airline crew paring, VLSI design, optimization for machine learning problems, drug discovery, etc \cite{CombApp1,CombApp2,RingOscApp}. However, because of the NP-Hard nature of most of these problems, solving them accurately on a conventional classical digital computing unit consumes exponential time or resources. Classical approximate algorithms like Goemans-Williamson's (GW) algorithm can be implemented on such conventional computers and can solve these problems approximately. But the GW algorithm, when used for the Max-Cut problem (a popular NP-Hard combinatorial optimization problem that we explore here), can only offer an approximation ratio (AR) guarantee  $\approx$ 0.878 or higher \cite{GW1}. Also, time complexity, or time to solution (TTS), for the GW algorithm has been found to still grow polynomially with graph size (though not exponentially) \cite{GWvsSAvsCIM}.

Various emerging computing schemes have been proposed and experimentally demonstrated recently, targeted to solve these problems more efficiently. Among them,  computing schemes that convert the problem to its corresponding Ising Hamiltonian \cite{NPCompleteIsingFormulation} and then find the ground state of the Hamiltonian are known as Ising computing schemes \cite{IsingComputingReview}. Quadratic Unconstrained Binary Optimization (QUBO) is also often used for the same purpose, and it has been shown that QUBO and the Ising model are equivalent via a linear transformation of variables \cite{Spin_Akerman1}.  

Ising computing has been implemented through various emerging technologies \cite{IsingComputingReview,GWvsSAvsCIM,CoherentIMvsQA}. Out of them, two popular classes of technologies are:

\begin{itemize}
    \item quantum annealing and its gate-based counterpart, meant for Noisy Intermediate Scale Quantum (NISQ) computing: Quantum Approximate Optimization Algorithm (QAOA) \cite{NISQReview,QAOA_Farhi} 
    \item phase-binarized oscillator (PBO) systems \cite{OscillatorIMReview}, also known as Oscillatory Neural Networks (ONN) \cite{SKONN_NCE}
\end{itemize}

A generalized mathematical model for PBOs, known as Kuramoto model \cite{Kuramoto}, has been used to propose  Ising computing under sub-harmonic injection locking (SHIL) in reports by Wang \textit{et al} \cite{Jaijeet1,Jaijeet2}. Since the Kuramoto model is highly device-physics-agnostic and has been shown to describe the phase dynamics of various kinds of oscillator devices, experimental demonstrations of PBO-based Ising computing have been reported using a wide range of device technologies: electronic inductor-capacitor (LC) oscillator \cite{Jaijeet3}, electronic ring oscillator \cite{RingOscApp,ChrisKim1,ChrisKim2}, electronic Schmitt Trigger circuit \cite{NikhilShukla1,SKONN_NCE}, metal-insulator-transition device \cite{MIT1,MIT2}, etc.. 

Spintronic oscillators have also been proposed recently for Ising computing \cite{Spin_Akerman1,Spin_Akerman2,Spin_Finocchio1,Spin_Finocchio2} since they offer various advantages in this regard. They operate at room temperature unlike quantum methods mentioned above, which need milli-Kelvins for operation \cite{NISQReview}. They can be coupled through various means like dipole coupling \cite{Spin_DB1,Spin_DB2,Spin_Ashwin}, spin-wave coupling \cite{Spin_Akerman1,Spin_Akerman2}, and electrical coupling \cite{Spin_Liu,Spin_Hyunsoo}. They operate at a very high frequency (in GHz) and consume very low energy.

\subsection{Our Contributions}

\begin{figure*}
\centering
\includegraphics[width=0.99\textwidth]{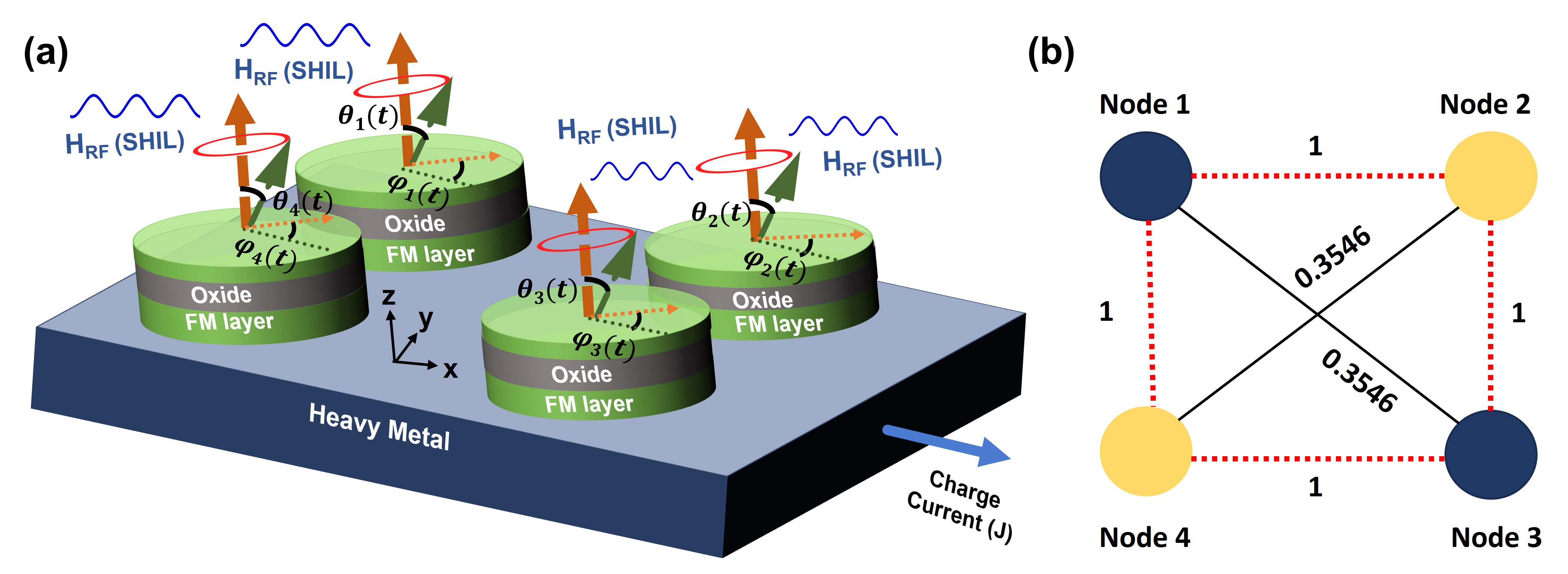}
\caption{(a) An array of dipole-coupled  SHNOs with each SHNO on the vertex of a square. The dynamics of each SHNO is given by the time evolution of its macro-spin vector, represented by $\theta(t)$ and $\phi(t)$ in spherical coordinates as shown. Uniform RF magnetic field $H_{RF}$ is applied to induce sub-harmonic injection locking (SHIL) (b) The weighted complete graph corresponding to the array in (a), with each edge weight value proportional to the dipole field between the two oscillators in (a), corresponding to the two nodes for the edge. Partitioning corresponding to the Max-Cut solution is also shown: nodes $1$ and $3$ belong to the first partition (blue circles), nodes $2$ and $4$ belong to the second partition (yellow circles), and edges connecting nodes of opposite partitions which contribute to the Max-Cut score (4 here) are shown as red dashed lines.} 
\label{SHNOArraySquareSchematic}
\end{figure*}

In the above context, we make the following contributions through this paper (we also show the section organization of the paper below): 

\begin{itemize}
    \item In Section 2, we model an array of four dipole-coupled uniform-mode spin Hall nano oscillators (SHNOs), as shown in Fig. ~\ref{SHNOArraySquareSchematic}(a), through Landau Lifshitz Gilbert Slonczweski (LLGS) equations, as well as through a more approximate model proposed by Slavin and Tiberkevich \cite{Slavin} (referred to as Slavin's model throughout this paper). We map the SHNO array to a complete weighted graph (Fig. ~\ref{SHNOArraySquareSchematic}(b)) \cite{Ref_CompleteGraph} and show that both LLGS model and Slavin's model of spin oscillators yield the same Max-Cut solution as Kuramoto model of generalized PBOs (which is independent of underlying device technology) and a classical brute force solver on the same weighted graph. This shows that spin oscillators follow the Kuramoto model of PBOs and can be used for solving the Max-Cut problem.

    \item In Section 3, restricting ourselves to the Max-Cut problem, 
    we solve for four kinds of graphs (unweighted Mobius Ladder \cite{Ref_MobiusLadder}, unweighted Random Cubic \cite{Ref_RandomCubic}, unweighted Erd{\"o}s R{\'e}nyi\cite{Ref_ErdosRenyi}, and weighted complete graphs \cite{Ref_CompleteGraph}) up to 20 graph nodes by numerically simulating a system of generalized PBOs, represented through Kuramoto model, and a system of spin oscillators represented through Slavin's model. We show that approximation ratio (AR), success probability (SP), and time to solution (TTS) show similar trends in both cases. We compare these results with two other alternative methods to solve the same MaxCut problem: classical GW algorithm, run on a conventional computer, and QAOA, meant to be run on gate-based NISQ hardware \cite{NISQReview,QAOA_Farhi}. We show that AR and SP obtained for generalized PBOs, as well as spin oscillators, are comparable to that for GW and much higher than that for QAOA for almost all graph instances. Moreover, unlike GW, TTS for generalized PBOs and spin oscillators does not grow with graph size for the instances we have studied.
\end{itemize}

\section{Modelling of an Array of Dipole-Coupled Spin Hall \\ Nano-Oscillators (SHNOs)}

\subsection{Landau Lifshitz Gilbert Slonczweski (LLGS) equations to model SHNOs}

Modeling of synchronization in a dipole-coupled heavy metal/ ferromagnetic metal (FM)/ oxide \\ heterostructure-based SHNO array, with four SHNOs located at four vertices of a square (as shown in Fig. ~\ref{SHNOArraySquareSchematic}(a)), has been discussed in detail in the context of neuromorphic computing in reports by Garg \textit{et al} \cite{Spin_DB1} and Hemadri Bhotla \textit{et al} \cite{Spin_DB2}. Each SHNO is assumed to work in the uniform mode. So, its magnetization dynamics are described by a single macrospin vector that represents all the magnetic moments of the ferromagnetic metal layer. Coupled equations that describe the phase dynamics of the SHNOs in spherical coordinates (for $n$-th SHNO, $\theta_n(t)$: polar angle of the macro spin vector for that SHNO, $\phi_n(t)$: azimuthal angle of the macro spin vector for that SHNO),  have been derived in \cite{Spin_DB2} (using LLGS equations that model spin-orbit-torque-driven magnetization dynamics) and are as follows: 

\begin{multline}
(1+\alpha^2)\frac{d\theta_n(t)}{dt} =  - \gamma \alpha H_{dipole}^{n,x}(t) sin(\theta_n(t)) 
- \gamma \beta \epsilon_n (cos (\theta_n(t)) cos (\phi_n(t)) + \alpha sin (\phi_n(t))) \\
+ \gamma H_{dipole}^{n,x}(t) (\alpha cos(\theta_n(t)) cos(\phi_n(t)) - sin(\phi_n(t))) \\
+ \gamma H_{dipole}^{n,y}(t) (\alpha cos(\theta_n(t)) sin(\phi_n(t)) +cos(\phi_n(t)))
\label{LLG_Coupled1}
\end{multline}

\begin{multline}
sin(\theta_n(t)) (1+\alpha^2)\frac{d\phi_n(t)}{dt} =  \gamma H_{dipole}^{n,z}(t) sin(\theta_n(t)) 
+ \gamma \beta \epsilon_n (sin(\phi_n(t)) - \alpha  cos (\theta_n(t)) cos (\phi_n(t))) \\
- \gamma H_{dipole}^{n,x}(t) (cos(\theta_n(t)) cos(\phi_n(t)) + \alpha sin(\phi_n(t))) \\
+ \gamma H_{dipole}^{n,y}(t) (\alpha cos(\phi_n(t)) - cos(\theta_n(t)) sin(\phi_n(t)))
\label{LLG_Coupled2}
\end{multline}

, where ($H_{eff}^{n,x}(t)\hat{x}+H_{eff}^{n,y}(t)\hat{y}+ H_{eff}^{n,z}(t)\hat{z}$), or $\vec{H_{eff}^{n}}$, is the net effective magnetic field experienced by SHNO labelled $n$. $\epsilon_n=\frac{\Lambda^2}{(\Lambda^2 + 1)+(\Lambda^2 - 1)sin(\theta_n(t))cos(\phi_n(t))}$, and $\beta= \frac{J\theta_{SHE}\hbar}{M_{sat}ed}$ \cite{Taniguchi}.  Here, $\gamma =$  gyromagnetic ratio, $\alpha =$ Gilbert damping parameter, $\theta_{SHE} =$ spin Hall angle of the heavy metal, $d=$ thickness of the ferromagnetic layer, $\Lambda =$ Slonczweski parameter, $M_{sat}$ = saturation magnetization, and $J$: current density through the heavy metal layer which triggers magnetic oscillations in SHNOs through spin Hall effect (as shown in Fig.~\ref{SHNOArraySquareSchematic}(a)). It is to be noted that here $\phi_n(t)$ represents the phase of oscillator $n$. 
 
In the reports by Garg \textit{et al} \cite{Spin_DB1} and Hemadri Bhotla \textit{et al} \cite{Spin_DB2}, $\vec{H_{eff}^{n}}$ contains contributions from the effective field due to perpendicular magnetic anisotropy (PMA) $H_k$, DC magnetic field applied along the out-of-plane direction $H_{DC}$, and RF dipole field due to the precessing macro-spin vectors of SHNOs (which couples the phase dynamics of $n$-th SHNO with other SHNOs).

For Ising computing, phase binarization under SHIL is an added property that's needed in these oscillators, as mentioned earlier. SHIL is carried out here by applying, on all SHNOs, a uniform RF magnetic field $H_{RF}$ of frequency twice the natural frequency of a SHNO, as shown in Fig. ~\ref{SHNOArraySquareSchematic} (a) \cite{Jaijeet1,Jaijeet2}. So, this uniform RF field also contributes to $\vec{H_{eff}^{n}}$ in our present model, along with PMA, applied DC field, and dipole RF fields due to other SHNOs mentioned above. The complete expressions for $\vec{H_{eff}^{n}}$ including all these four factors, for $n=$ 1, 2, 3, 4 in the case of four SHNOs on four vertices of a square (Fig. ~\ref{SHNOArraySquareSchematic}(a)), are provided in Section 1 of Supplementary Information accompanying this paper.

Using these expressions for $\vec{H_{eff}^{n}}$, equations ~\ref{LLG_Coupled1} and ~\ref{LLG_Coupled2} are solved numerically for this work. The following values are used for the parameters in the equations: 
$M_{sat}$ = $1.313\times$ 10$^6$ A/m, $H_k$ = 1.79 T, $H_{DC}$ = 0.1 T, $d$ = 2 nm, $\gamma$ = $17.6 \times 10^{10}$ $Hz/T$, $\alpha$ = 0.005, $\theta_{SHE}$  = 0.07, and $\Lambda$  = 2 \cite{Spin_DB1,Spin_DB2,Taniguchi}. These values are chosen based on the experimental study of magnetic tunnel junction (MTJ) devices and experimental characterization of spin accumulation at heavy metal- ferromagnetic metal interface due to in-plane current flow  \cite{Taniguchi_Expt1,Taniguchi_Expt2, Taniguchi_Expt3, SHE_Pt_1,SHE_Pt_2}.

As shown in Fig. ~\ref{SHNOArraySquareSchematic}(a), all the SHNOs experience magnetic oscillations due to the same current density through the heavy metal layer: $J = 1 \times 10^{11}$ A/m$^2$ and hence have the same natural frequency of 3.77 GHz. So, an external RF field of frequency 7.54 GHz (twice the natural frequency) is applied for SHIL  in accordance with the SHIL scheme proposed by Wang \textit{et al} \cite{Jaijeet1,Jaijeet2}. The center-to-center distance between two adjacent SHNOs in a square array (Fig. ~\ref{SHNOArraySquareSchematic}(a)) is considered to be 225 nm, and hence the distance between two SHNOs on the diagonal is 318 nm.

\subsection{Slavin's Model of Spin Oscillators}

In the model proposed by Slavin and Tiberkevich \cite{Slavin} for spintronic auto-oscillators, each oscillator $j$ is represented by its complex spin-oscillation amplitude $c_j$ and dimensionless spin wave power $p_j=|c_{j}^2|$ which is proportional to the experimentally measured microwave power of the spin oscillator. The equations used to model the array of coupled spin oscillators, as per this Slavin's model, are given by:

\begin{equation}
    \frac{d c_j}{d t}=-i \omega_j\left(p_j\right) c_j-\Gamma_{+, j}\left(p_j\right) c_j+\Gamma_{-, j}\left(p_j\right) c_j +
    K_e e^{-i \omega_e t} c_j^*+\sum_{j^{\prime}} \Omega_{j, j^{\prime}} e^{i \beta_{j, j}} c_{j^{\prime}}
    \label{Slavin1}
\end{equation}

Equation ~\ref{Slavin1} can be decoupled into two equations for power $p_j(t)$ and phase $\hat{\phi}_j(t)$ of $c_j$ as:

\begin{equation}
    \frac{d p_j}{d t}=-2 p_j\left[\Gamma_{+, j}\left(p_j\right)-\Gamma_{-, j}\left(p_j\right)\right] 
    +2 K_e p_j \cos \left(\omega_e t+ 2 \hat{\phi}_j \right) +  
    2 \sum_{j, j^{\prime}} \Omega_{j, j^{\prime}} \sqrt{p_j p_{j^{\prime}}} \cos \left(\hat{\phi}_j-\hat{\phi}_{j^{\prime}}-\beta_{j, j^{\prime}}\right)
\label{Slavin2}
\end{equation}

\begin{equation}
\frac{d\hat{\phi}_j}{d t}=-\omega_j\left(p_j\right)-K_e \sin \left(\omega_e t+2 \hat{\phi}_j\right) \\
+\sum_{j, j^{\prime}} \Omega_{j, j^{\prime}} \sqrt{p_{j^{\prime}} / p_j} \sin \left(\hat{\phi}_{j^{\prime}}-\hat{\phi}_j+\beta_{j, j^{\prime}}\right)
\label{Slavin3}
\end{equation}

, where $\omega_j\left(p_j\right)= \omega_0+\mathcal{N}p_j$ is the precession term which represents the operating frequency, $\mathcal{N}/2\pi\approx -3.44$ GHz is the nonlinear frequency shift coefficient, $\Gamma_{+, j}\left(p_j\right)=\Gamma_{G}(1+Qp_j)$ is the damping term with $Q\approx2.66$ and $\Gamma_{G}/2\pi\approx 252$ MHz, $\Gamma_{-, j}\left(p_j\right)=\sigma I_{dc}(1-p_j)$ is the anti-damping term due to the spin polarised direct current $I_{dc}$ and $\sigma \approx 2502.1998$ GHz/A. The above parameter values are taken from the report by Albertsson \textit{et al} \cite{Spin_Akerman3}. Here, $\Omega_{j, j^{\prime}}$: coupling constant between oscillators $j$ and $j^{\prime}$, $\beta_{j, j^{\prime}}$: phase constant between oscillators $j$ and $j^{\prime}$. For SHIL, an external RF current of frequency $\omega_e$ and coupling strength $K_e$  is injected into the spin oscillators. 

In our work, we solve the coupled differential equations ~\ref{Slavin2} and ~\ref{Slavin3} numerically, with time range from 0 to 200 ns divided into $100,000$ intervals of equal size. We use $\beta_{j, j^{\prime}}=-1.6\pi$, $K_e=25$ MHz, $\omega_0/2\pi=4.2$ GHz, and $I_{dc}=1.5$ mA. Corresponding to two adjacent oscillators $j$ and $j'$ in the oscillator array of Fig. ~\ref{SHNOArraySquareSchematic}(a), we use $\Omega_{j, j^{\prime}}$ = 25 MHz in our numerical model. Corresponding to oscillators $j$ and $j'$ on two ends of a diagonal of the square (Fig. ~\ref{SHNOArraySquareSchematic}(a)), we use $\Omega_{j, j^{\prime}}$ = 8.865  MHz. These values are chosen in accordance with the fact that the dipole field decays as 1/$R^3$ if $R$ is the distance between two SHNOs (as shown in reports by Garg \textit{et al} \cite{Spin_DB1}, Hemadri Bhotla \textit{et al} \cite{Spin_DB2}, and Amin \textit{et al} \cite{Amin}, and also in Section 1 of Supplementary Information).


\subsection{Kuramoto Model for Generalized PBOs}

According to the Kuramoto model (used here to model generalized PBOs, independent of the exact device physics), the net energy of a system of coupled oscillators, under sub-harmonic injection locking (SHIL), is given by:

\begin{equation}
E=-\Sigma_{j'=1}^{n}\Sigma_{j=1,j < j'}^{n}K_{j,j'}\cos(\overline{\phi}_{j}-\overline{\phi}_{j'}) 
-K_{s}\sum_{j=1}^{n}\cos (2\overline{\phi}_{j})
\label{Kuramoto1}
\end{equation}

, where $\overline{\phi}_j$: phase of oscillator $j$, $K_s$: anisotropy constant (leads to SHIL), and $K_{j,j'}$: coupling constant between oscillator $j$ and oscillator $j'$ \cite{Spin_Liu}.


Phases of oscillators evolve over time ($t$) as follows:

\begin{equation}
\frac{d\overline{\phi}_j}{dt}=\omega _{j}-\omega^{*} - \sum _{j'(j'\ne j)}^{N}K_{j,j'}\sin (\overline{\phi}_{j}-\overline{\phi}_{j'})-2K_{s}\sin (2\overline{\phi}_{j})
\label{Kuramoto2}
\end{equation}

, where $\omega_j$: natural frequency of oscillator $j$,   $\omega^{*}$: average natural frequency of all the oscillators \cite{Spin_Liu}. When the natural frequencies of the oscillators are equal (as in the cases considered in this paper), $\frac{d \overline{\phi}_j}{dt} = -\frac{\partial E}{\partial\overline{\phi}_j}$, and thus the phases evolve such that the net energy of the system $E$ (as per equation ~\ref{Kuramoto1}) goes to a minimum, as expected in Ising computing \cite{Spin_Liu}. 

We solve equation ~\ref{Kuramoto2} numerically using Runge Kutta method, with each time-step $\Delta t=5 \times 10^{-5}$ s, and number of time-steps =$10^{6}$. We use the following values in the Kuramoto model: $\omega_j$ =0.1 radians/s for all oscillators labelled $j$ from 1 to $N$ (hence, $\omega^{*}$ = 0.1 radians/s), and $K_{s}=0.5$ radians/s. In congruence with $1/R^3$ relationship of dipole field between oscillators, we choose $K_{j, j^{\prime}} = -1$ radian/s  when oscillators $j$ and $j'$ are adjacent in the square array, and $K_{j, j^{\prime}}=-0.354$ radians/s   when oscillators $j$ and $j'$ are on a diagonal of the square array (Figure ~\ref{SHNOArraySquareSchematic}(a)).

\begin{figure}
\centering
\includegraphics[height=0.8\textheight]{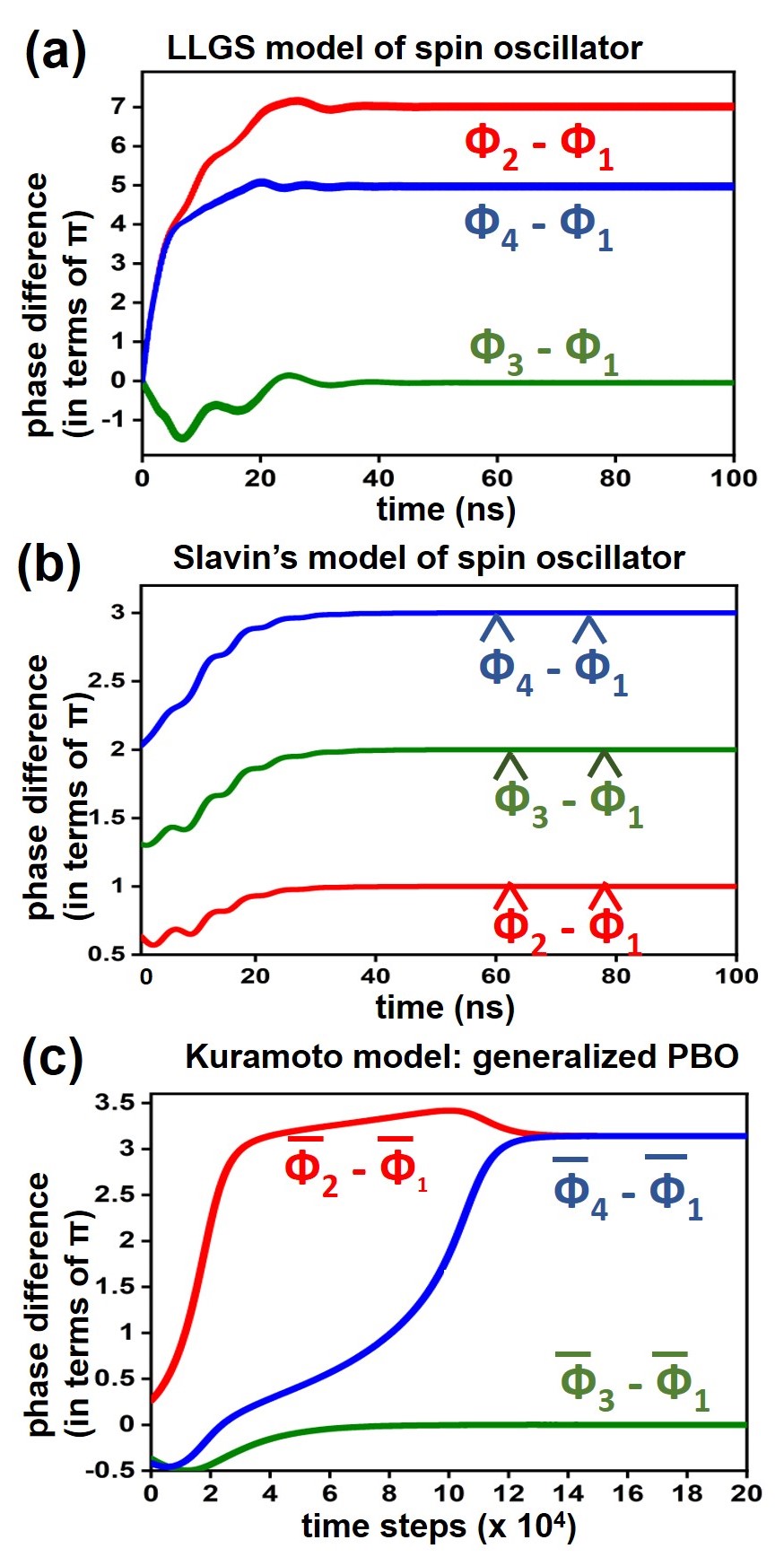}
\caption{For the oscillator array of Fig. 1(a), phase differences between oscillators $2$ and oscillators $1$ (red plot), oscillators $3$ and oscillators $1$ (green plot), and oscillator $4$ and oscillator $1$ (blue plot) are plotted here with respect to time as obtained from: (a) LLGS model of spin oscillator, (b) Slavin's model of spin oscillator, and (c) Kuramoto model for generalized PBO.} 
\label{SHNOArraySquarePhases}
\end{figure}

\subsection{Phase Dynamics of Four Oscillators: Comparing LLGS, Slavin, and Kuramoto models}

The phase dynamics of the four oscillators of the array in Fig. ~\ref{SHNOArraySquareSchematic}(a) is solved using LLGS, Slavin, and Kuramoto models as described above. The differences between phases of oscillators labeled 2, 3, and 4 and the phase of reference oscillator (labeled 1) are plotted in Fig. ~\ref{SHNOArraySquarePhases}(a),(b), (c) respectively. We observe that for all three models, the phase difference between oscillators $2$ and $1$, and oscillators $4$ and $1$ (adjacent oscillators) is an odd multiple of $\pi$. But the phase difference between oscillators $3$ and $1$ (oscillators on the diagonal) is an even multiple of $\pi$. 

Thus, with respect to the weighted complete graph corresponding to this oscillator array shown in Fig.~\ref{SHNOArraySquareSchematic}(b), all three oscillator models yield the following solution: nodes 1 and 3 are in one partition/ blue circles (since oscillators 1 and 3 have phase difference in even multiples of $\pi$, or effectively same phases), and node 2 and 4 are in the other partition/ yellow circles (since oscillators 1 and 2 and oscillators 1 and 4 have phase difference in odd multiples of $\pi$, or effectively opposite phases). The corresponding cut score is 4. This value matches with the Max-Cut score for the graph of Fig.~\ref{SHNOArraySquareSchematic}(b), as obtained from a classical brute force solver that we have developed for this purpose. But the brute force solver tries out all partitions one by one, and calculates the maximum cut score from them (Max-Cut score), and hence its time complexity (or TTS) grows exponentially with graph size. 

For weighted graphs like in Fig.~\ref{SHNOArraySquareSchematic}(b), the Max-Cut problem is about partitioning the given graph into two sets such that the sum of the weights of all the edges that connect nodes of one partition (blue circles) with nodes of the other partition (yellow circles) is maximum, and the corresponding sum of edge weights is the Max-Cut score for the graph. For unweighted graphs used in the next section, the problem reduces to maximizing the number of edges that connect nodes of one partition with nodes of the other partition, and the corresponding number of edges is the Max-Cut score. 

In Section 2 of Supplementary Information, we consider four dipole-coupled SHNOs arranged in a line/ chain, with adjacent SHNOs separated by a fixed distance. We show that even in that case, both LLGS and Slavin's model of SHNOs predict the same Max-Cut partitioning and Max-Cut score as the classical brute force solver acting on the weighted complete graph corresponding to this new configuration of SHNOs.

\section{Comparing Performance of SHNOs with GW algorithm and QAOA}

\subsection{Slavin's Model, Kuramoto Model, and Solving Max-Cut on Four Types of Graphs}

Having shown in the previous section that spin oscillators yield the correct Max-Cut solution for small graphs, here we solve much larger size graphs by numerically solving the Kuramoto model of generalized PBOs and Slavin's model of spin oscillators, with the number of oscillators being equal to the number of nodes of the graph and oscillator-to-oscillator coupling coefficients chosen based on edge connectivity/ weights of edges of the graph. As mentioned in Section 1, we have chosen four kinds of graphs for this purpose: unweighted Mobius Ladder \cite{Ref_MobiusLadder}, unweighted Random Cubic \cite{Ref_RandomCubic}, unweighted Erd{\"o}s R{\'e}nyi\cite{Ref_ErdosRenyi}, and weighted complete graphs \cite{Ref_CompleteGraph}.  All these graphs have been defined formally in Section 3 of Supplementary Information, and some instances of these graphs have been shown there. For Erd{\"o}s R{\'e}nyi graphs, we choose an edge connectivity probability of 0.5 because it corresponds to the highest difficulty level in terms of solving the Max-Cut problem. Graphs with both very low or very high levels of connectivity have been found to be much easier problem instances compared to that \cite{QAOA_ErdosRenyi}.

For unweighted graphs, we take  $K_{j, j^{\prime}}=1$ radan/s in Kuramoto model (equation ~\ref{Kuramoto2}) and $\Omega_{j, j^{\prime}}=25$ MHz (equation ~\ref{Slavin2}, ~\ref{Slavin3}) in Slavin's model corresponding to all existing edges between node $j$ and node $j^{\prime}$, and  $K_{j, j^{\prime}}=0$ in Kuramoto model and $\Omega_{j, j^{\prime}}=0$  in Slavin's model whenever no edge exists between nodes labeled $j$ and $j^{\prime}$. For weighted graphs, we take $K_{j, j^{\prime}}=w_{j, j^{\prime}}$ radian/s in Kuramoto model and $\Omega_{j, j^{\prime}}=w_{j, j^{\prime}}\times25$ MHz where the weight between node $j$ and node $j'$ is $w_{j, j^{\prime}}$. 

We choose Slavin's model instead of the LLGS model to model spin oscillators for larger graph problems because, for Slavin's model, coupling coefficients between oscillators can be chosen arbitrarily. Hence, any graph configuration can be mapped to the oscillators. On the other hand, in the LLGS model, the coupling coefficient is determined by a physical phenomenon (dipole coupling in our case) and hence can't be chosen arbitrarily; only a subset of all possible graphs can be mapped to dipole-coupled SHNO arrays (like the complete weighted graph shown in Fig. ~\ref{SHNOArraySquareSchematic}(b) and that shown in Fig. 2(b) of Supplementary Information). Using electrical coupling instead of dipole coupling offers more flexibility in this mapping \cite{Spin_Liu,Spin_Hyunsoo} . Modeling electrically coupled SHNOs for the different graph instances used in this section will be a subject of our future study. 

For each graph instance, we solve equations ~\ref{Slavin2} and ~\ref{Slavin3} numerically 100 times (100 trials) for Slavin's model and solve equation ~\ref{Kuramoto2} numerically 100 times (100 trials) for the Kuramoto model, with different initial phase values each time, to obtain the phase dynamics of the oscillators and their steady-state phase solutions. From each steady-state phase solution, the corresponding graph partitioning and cut score are obtained just like in Section 2, based on the difference between the phase of each oscillator and that of the reference oscillator (oscillator 1 in Section 2). Thus, for each graph instance, we obtain 100 cut values for the Kuramoto model and 100 cut values for Slavin's model, from which AR and SP are calculated as described later. 


\begin{figure*}[!t]
     \centering
\includegraphics[width=0.99\textwidth]{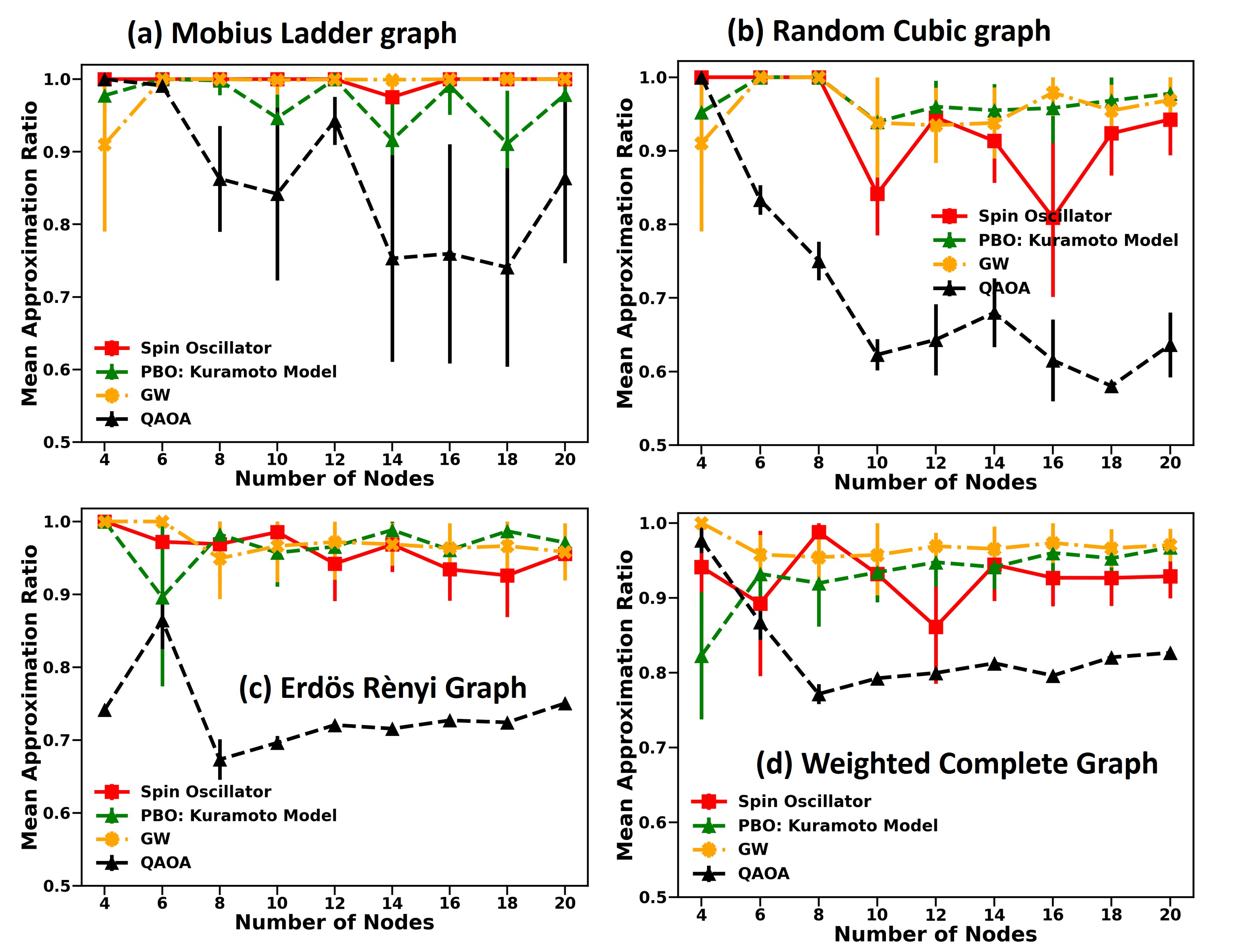}
\caption{Approximation ratio (AR), averaged across a large number of trials/ cut results, as a function of number of nodes of the graph, for Slavin's model of spin oscillators (red plot), Kuramoto model of generalized PBOs (blue plot), classical GW algorithm (orange), and QAOA (black). Four kinds of graphs are used: (a) Mobius ladder, unweighted, (b) random cubic, unweighted, (c) Erd{\"o}s R{\'e}nyi, unweighted, (d) complete graph, weighted. Error bars correspond to standard deviation across the same number of trials/ cut results.}
\label{ApproxRatio}
\end{figure*}

\begin{figure*}[!t]
     \centering
\includegraphics[width=0.99\textwidth]{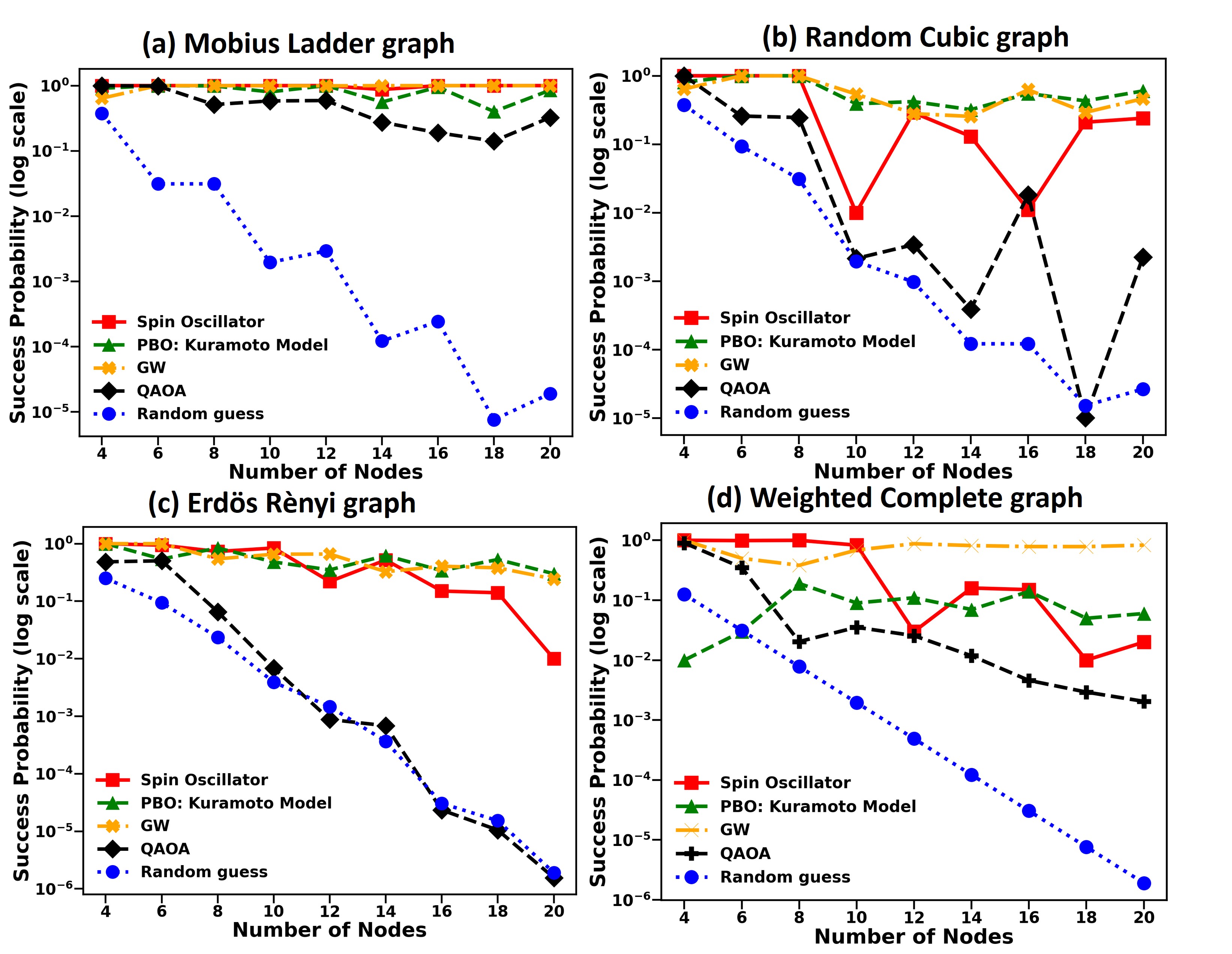}
\caption{Success probability (SP), over a large number of trials/ cut results, as a function of number of nodes of the graph, for Slavin's model of spin oscillators (red plot), Kuramoto model of generalized PBOs (blue plot), classical GW algorithm (orange), and QAOA (black) for the four kinds of graph considered in this work.}
\label{SuccessProb}
\end{figure*}

\subsection{Classical Goemans-Williamson's (GW) algorithm}

Given an undirected weighted graph $\mathcal{G}=(V,E)$, where $V$ and $E$ are the graph's vertex and edge sets respectively, and weights $w_{ij}$, the objective function for Max-Cut can be expressed as 
\begin{equation}
\label{eqn:maxcut_def_orig}
\begin{split}
        \text{max}& \hspace{20pt} \frac{1}{2}\sum\limits_{(i,j)\in E}w_{ij}(1-x_i x_j)\\
        \text{subject to}& \hspace{20pt} x_i\in\{-1,1\} \hspace{22pt},\text{ for} \hspace{5pt} i \in V
\end{split}
\end{equation}
For unweighted graphs, $w_{ij}=1$ when an edge connects node $i$ and $j$ in the graph, otherwise 0.

Classical GW algorithm 
optimizes the above objective function through appromixation (as described below). Since GW algorithm can be implemented on conventional digital computers, it is 
very commonly used for solving the Max-Cut problem on large graphs \cite{GW1}.
In the GW formulation, the MaxCut objective function is relaxed and reformulated as: 
\begin{equation}
\begin{split}
       \text{max}& \hspace{20pt} \frac{1}{2}\sum\limits_{(i,j)\in E}w_{ij}(1-X_{ij}) \\
       \text{subject to}& \hspace{20pt} \Vec{a}^T X \Vec{a}\ge 0 \hspace{10pt},\text{ } \forall \Vec{a}\in \mathbb{R}^N\\
       &\hspace{20pt} X_{ii}=1 \hspace{22pt},\text{ for} \hspace{5pt} i \in V\\
       &\hspace{20pt}  X\in \mathbb{R}^{N\times N}\\
\end{split}
\end{equation}
, where $X_{i,j}=\Vec{x}_i\cdot\Vec{x}_j$, $(\cdot)$ is the inner product and $|V|=N$ (size of the graph). The binary variables $x$ in equation (\ref{eqn:maxcut_def_orig}) are replaced by vectors $\Vec{x}$ on an $N$-dimensional unit sphere i.e. $\Vec{x}\in\mathbb{S}^N$. This is known as semi-definite programming (SDP) relaxation. Solving for $X$ yields the optimal vector $\Vec{x}_{\text{opt}}$ for every vertex. We have used here the CVXOPT \cite{andersen2013cvxopt} and SciPy \cite{2020SciPy-NMeth} python packages to solve for $\Vec{x}_{\text{opt}}$.

Next, we draw a random hyperplane passing through the origin of $\mathbb{S}^N$ with the corresponding normal vector to the plane being $\Vec{r}$. The rationale behind this step is to partition the vectors into two sub-spaces where vectors in one subspace are assigned the value $-1$ and in the other sub-space $1$, after investigating the sign of $(\Vec{x}_{\text{opt}}\cdot\vec{r})$. If this inner product is positive (negative), then $\Vec{x}_{\text{opt}}$ is mapped to $+1 (-1)$. This part of the algorithm is the measurement step and is explained with a schematic in Section 4 of the Supplementary Information.  We take $1024$ such random normal vectors $\Vec{r}$ and thus obtain 1024 cut values, from which AR and SP are obtained.


\subsection{Quantum Approximate Optimization Algorithm (QAOA)}

QAOA is a popular way of solving the Max-Cut problem on NISQ-era quantum hardware \cite{QAOA_Farhi,NISQReview,QuantumAnnealingvsQAOA1}. Here, we briefly describe the steps in QAOA, in relation to solving MaxCut. For this work, we have implemented QAOA on the Qiskit quantum simulator \cite{Qiskit}, without incorporating any noise model in it. Details of our numerical method and a schematic of the QAOA circuit we have used are presented in Section 5 of Supplementary Information.

The MaxCut problem of any graph of $N$ nodes is solved through QAOA in the Hilbert space of $2^{N}$ dimensions, with each basis state corresponding to a possible partitioning of the graph. Thus solving this problem through a quantum circuit involves $N$ qubits. The given graph $(V,E)$ is converted to the corresponding problem Hamiltonian ($H_{P}$) in the same Hilbert space as follows \cite{QAOA_Farhi, QAOA_Tutorial}:

\begin{equation}
H_P= \Sigma_{(j,j^{'})\in E}(\frac{1}{2}w_{j,j'}((I^{1}
\otimes I^{2} \otimes ... I^{j}\otimes...I^{j'}\otimes...I^{N})
-(I^{1} \otimes I^{2} \otimes .... \sigma_z^{j}\otimes ...\sigma_z^{j'}\otimes... I^{N})))
\label{QAOA_Hp}
\end{equation} 

, where $\sigma_z$: Pauli spin matrix (in z), $I$: identity matrix, $w_{j,j'}$: weight of an edge connecting node $j$ with node $j'$ for the corresponding weighted graph (for unweighted graph, $w_{j,j'}=1$ if an edge exists between node $j$ and $j'$, and 0 otherwise). 

As the first step of the forward pass of every iteration, the initial state $\ket{s}$ is freshly prepared by applying the Hadamard gate on all the $N$ qubits initialized to $\ket{0}$ each (Fig. 6 of Supplementary Information) \cite{QAOA_Tutorial}. Then,  we let the state $\ket{s}$ evolve through multiple stages of unitary operators involving mixing Hamiltonian ($H_M$) and problem Hamiltonian ($H_P$) (defined above) and coefficients/ learning parameters ($\beta_1$, $\gamma_1$, $\beta_2$, $\gamma_2$, ... $\beta_p$, $\gamma_p$; if there are total $p$ stages), as follows \cite{QAOA_Farhi, QAOA_Tutorial}: 

\begin{equation}
    \ket{\psi}=e^{-i\beta_pH_M}e^{-i\gamma_pH_P}.....
    e^{-i\beta_2H_M}e^{-i\gamma_2H_P}e^{-i\beta_1H_M}e^{-i\gamma_1H_P} \ket{s}
    \label{QAOA_evol}
\end{equation}.

This is the set of unitary operations corresponding to the forward-pass quantum circuit here. The mixing Hamiltonian $H_M$ is given by: 

\begin{equation}
H_M=\sum _{i=1}^{N}(I^{1}\otimes I^{2} \otimes I^{3}...I^{i-1} \otimes \sigma _{x}^{i}\otimes I^{i+1}....\otimes I^{n})
\label{QAOA_Hm}
\end{equation}.

The gate decomposition of the unitary operators $e^{-i\gamma H_P}$ and $e^{-i\beta H_M}$ into single-qubit and two-qubit gates is shown in Fig. 6 of Supplementary Information. 


The expectation value of the problem Hamiltonian for the evolved state ($\bra{\psi}H_P\ket{\psi}$) is maximized over several iterations, where each iteration consists of a forward pass following equation 2 and an update of ($\beta_1$, $\gamma_1$, $\beta_2$, $\gamma_2$, ... $\beta_p$, $\gamma_p$) following some classical optimization algorithm which makes use of the expectation value $\bra{\psi}H_P\ket{\psi}$ \cite{QAOA_Farhi,QAOA_Tutorial}. More details of our numerical method we have followed can be found in Section 5 of Supplementary Information.

With the final values of ($\beta_1$, $\gamma_1$, $\beta_2$, $\gamma_2$, ... $\beta_p$, $\gamma_p$), the forward pass and subsequent measurement (corresponding to $\sigma_z$ operator) are carried out 1024 times (when no success is obtained, number of measurements is increased further to 5,000,000 to obtain non-zero SP). For each measurement outcome, the corresponding graph partitioning is obtained as follows: qubit $i$ being $\ket{0}$ means the corresponding node $i$ belongs to the first partition, and qubit $i$ being $\ket{1}$ means the corresponding node $i$ belongs to the second partition. Based on the partitioning, the corresponding cut score is calculated like before. The iterative adjustment of ($\beta_1$, $\gamma_1$, $\beta_2$, $\gamma_2$, ... $\beta_p$, $\gamma_p$) starting from different initial values of these parameters is repeated 10 times here. Thus, we obtain 10 sets of 1024 cut values (or 5,000,000 sometimes) in the case of QAOA, from which AR and SP are obtained.

\begin{figure}[!t]
    \centering
\includegraphics[width=0.99\textwidth]{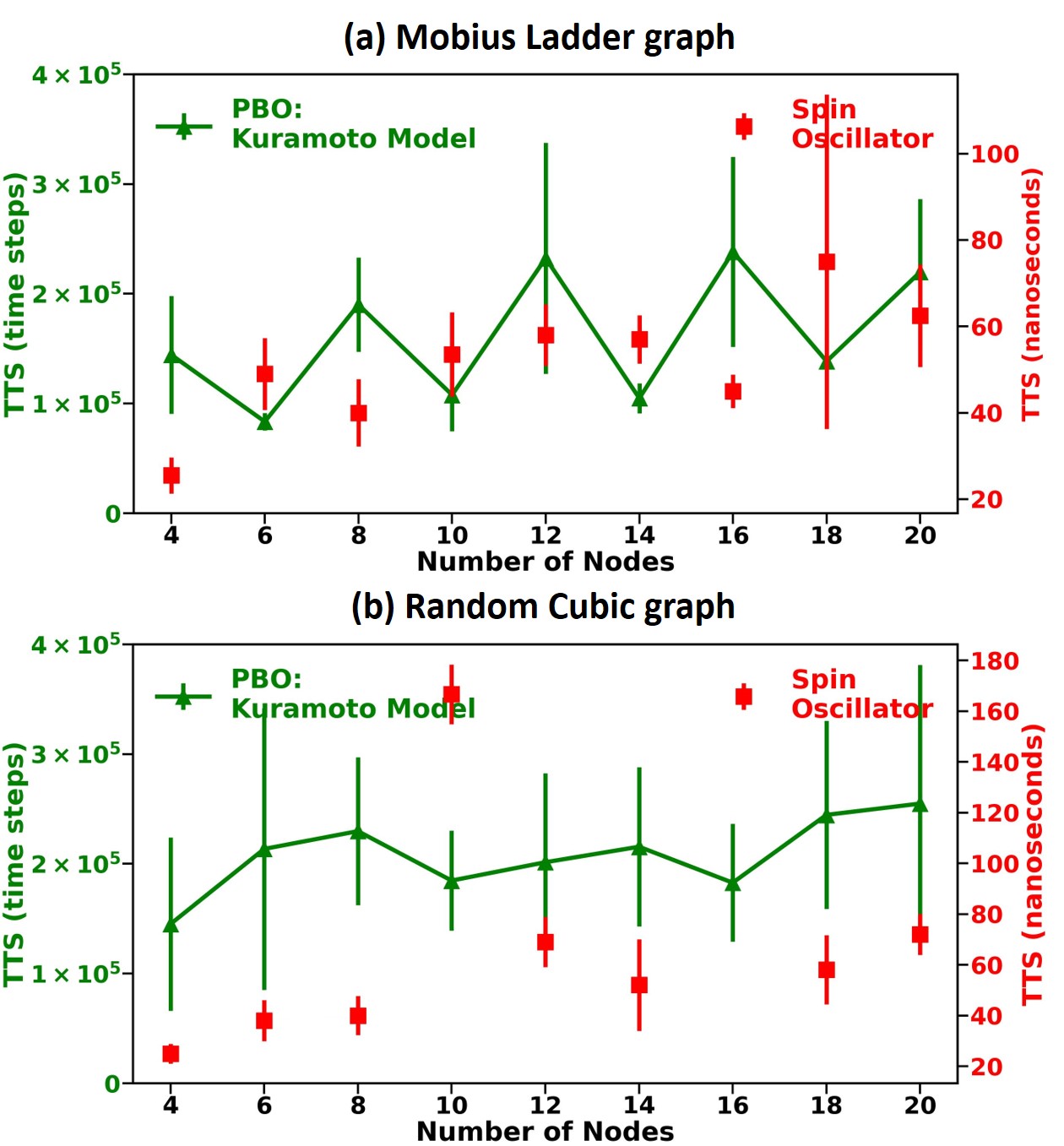}
\caption{Mean TTS (in number of time steps) for generalized PBOs, simulated using Kuramoto model, is compared against mean TTS (in nanoseconds) for spin oscillators, simulated using Slavin's model for (a) Mobius Ladder graph (b) random cubic graph. The mean value of TTS is calculated from 10 successful trials. Standard deviation across these trials is shown as an error bar.}
\label{TTS1}
\end{figure}

\begin{figure*}[!t]
    \centering
\includegraphics[width=0.99\textwidth]{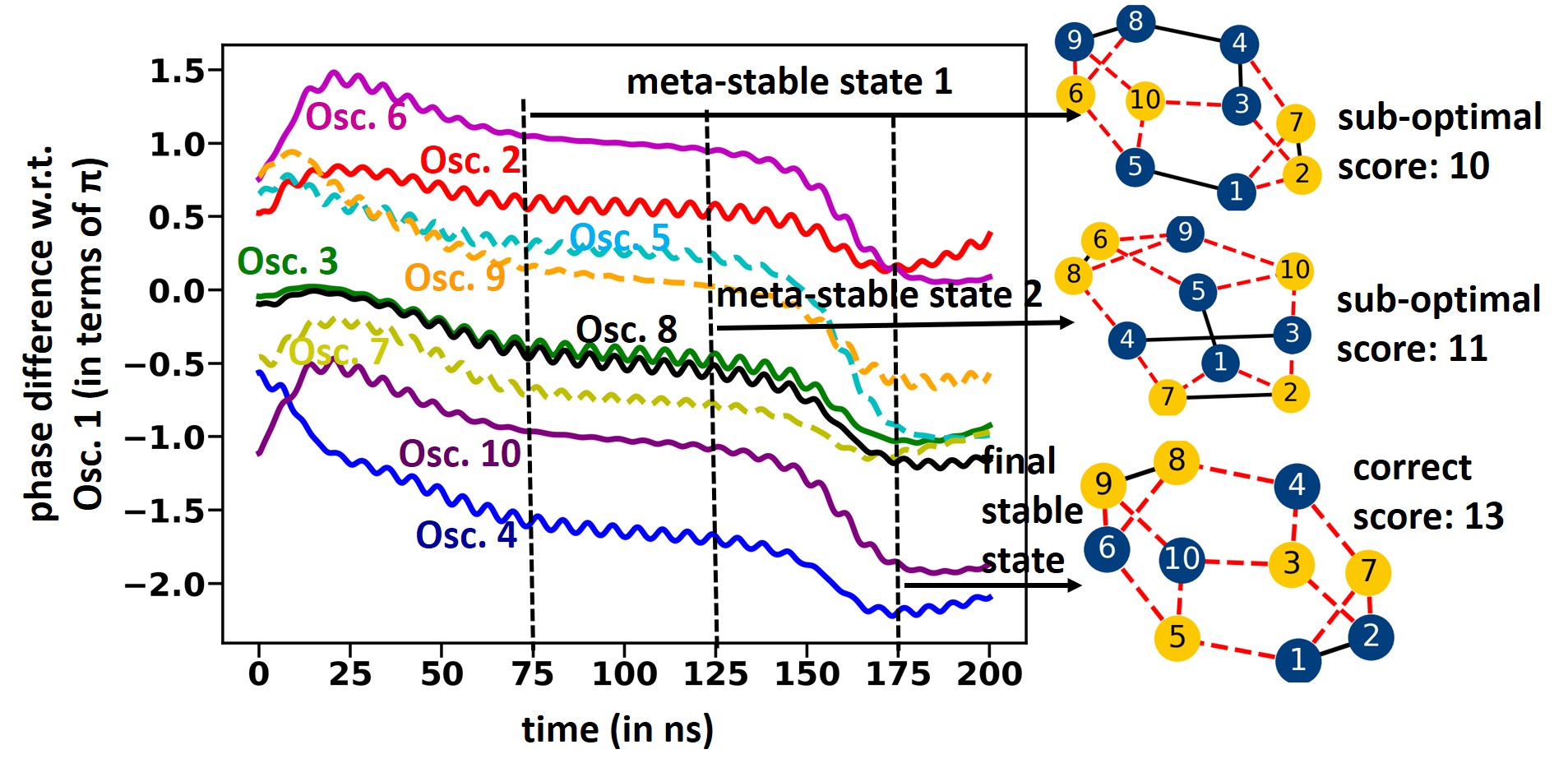}
\caption{Evolution of the phases of spin oscillators, modelled by Slavin's model, over time corresponding to the 10-node random cubic graph instance we have used in this work. Two meta-stable states with sub-optimal partitioning and the final state with optimal partitioning (correct Max-Cut score) are also shown. TTS here is 175 ns, as shown. (`w.r.t.': with respect to, `Osc.': Oscillator)}
\label{TTS2}
\end{figure*}

\subsection{Comparison Based on Approximation Ratio (AR) and Success Probability (SP)} 

For each graph instance, the average/ mean of the cut values obtained for a given model (Kuramoto model: 100, Slavin's model: 100, GW: 1024, QAOA: 10 $\times$ 1024, as mentioned above) is divided by the correct Max-Cut score obtained from a classical brute-force solver, applied on the same graph instance, to obtain the approximation ratio (AR). ARs for the four types of graphs used here, with graph sizes ranging from 4 to 20 nodes, are shown in Fig. ~\ref{ApproxRatio}.  Standard deviation across those cut values, divided by the correct Max-Cut score, is plotted as an error bar in these plots.

For each graph instance and each model, success probability (SP) is obtained as the ratio of the number of trials that yield the correct Max-Cut score to the total number of trials and plotted in Fig. ~\ref{SuccessProb}. For the obtained score from the given method to be called correct, it needs to match with that from classical brute-force solver exactly for unweighted graphs and be within 5\% error margin for unweighted graphs, as per our metric. To obtain finite SP for QAOA in some instances, total number of measurements for every run is increased to 5,000,000 as mentioned earlier, leading to higher number of trials for these instances compared to the others.

Fig. ~\ref{ApproxRatio} and Fig. ~\ref{SuccessProb} show that the AR and SP of spin oscillators closely match with that of generalized PBOs (Kuramoto model) for almost all types of graphs, which is expected from our finding for two specific 4-node weighted complete graphs in Section 2. ARs obtained both for spin oscillators and generalized PBOs are almost as high as that for classical GW algorithm for most types of graphs. This makes these oscillators very suitable for practical Ising computing solutions given that GW, implemented on conventional computers, is the popular method currently to solve Max-Cut for large graphs (as mentioned earlier) \cite{SKONN_NCE}. The classical brute-force solver which gives an exact solution and is used as the benchmark solver here cannot be used for such large graphs because its time complexity (or TTS) grows exponentially with graph size (as also mentioned earlier). 

Also, for all types of graphs, both AR and SP for spin oscillators and generalized PBOs are much better than those of QAOA. In fact for random cubic, Erd{\"o}s R{\'e}nyi, and weighted complete graphs, solving which is much more difficult than Mobius Ladder graphs (Mobius Ladder graphs, in fact, have polynomial-time solutions \cite{MobiusLadder_PolySoln}), SP of QAOA drops exponentially as the graph size increases while that of generalized PBOs and spin oscillators only drops linearly (or stays almost constant) with graph size. This shows that spin oscillators, and other PBOs in general, are much more attractive options for solving these kinds of combinatorial optimization problems compared to the NISQ-era gate-based quantum computing algorithm QAOA.

\begin{figure}
    \centering
\includegraphics[width=0.9\textwidth]{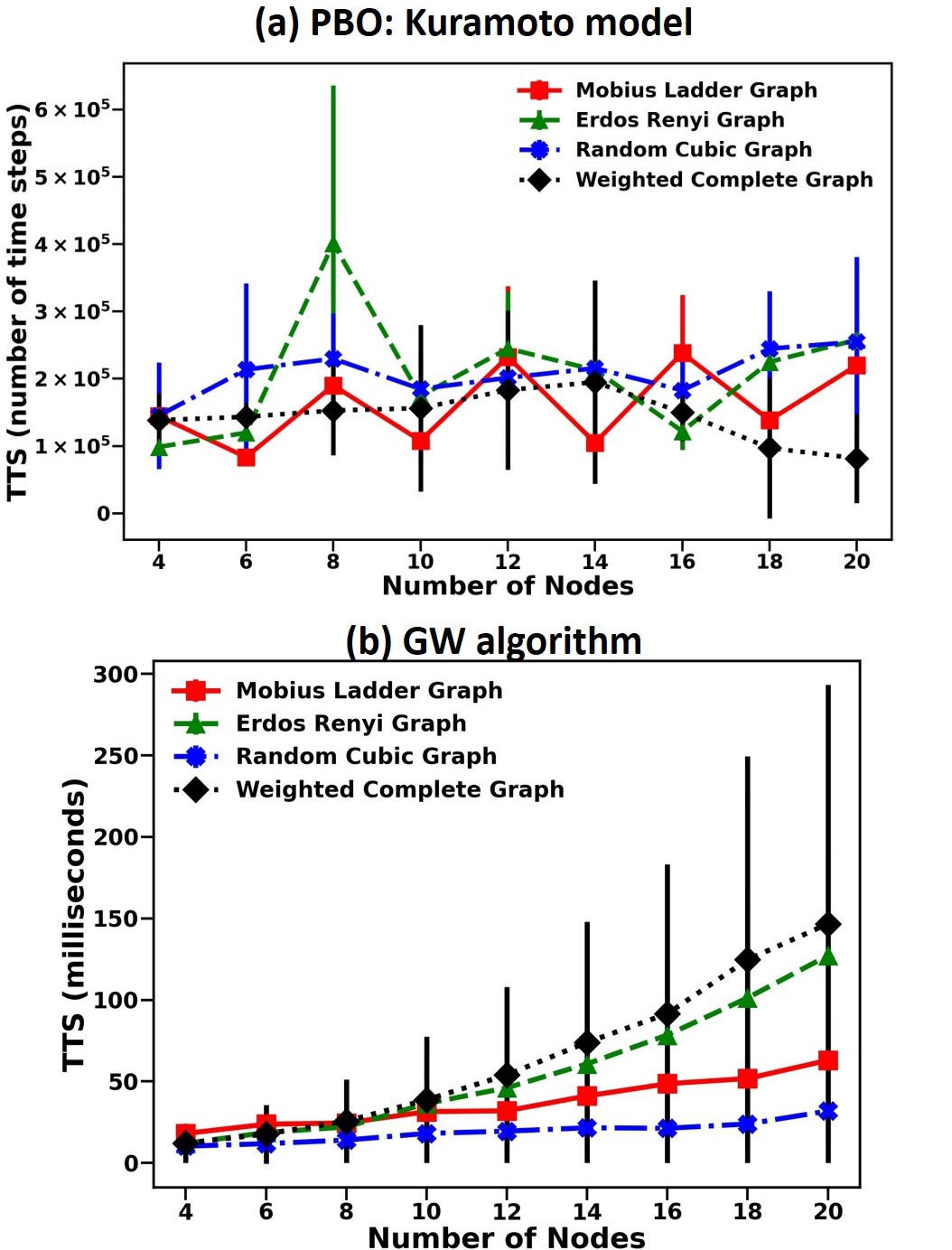}
\caption{(a) Mean TTS (in the number of time steps) for generalized PBOs, simulated using Kuramoto model, vs graph size for the four types of graph considered in this work. (b)  Mean TTS (in milliseconds) for classical GW algorithm vs graph size for the same four graph types. The standard deviation across trials in both cases is shown with an error bar.}
\label{TTS3}
\end{figure}

\subsection{Comparison Based on Time to Solution (TTS)}

To obtain time to solution (TTS) on a given graph instance for the Kuramoto model of generalized PBOs and Slavin's model of spin oscillators, we select 10 successful trials and calculate the average time and standard deviation of that  (in nanoseconds for spin oscillators, in number of time steps for generalized PBOs) when the phases of the oscillators reach steady state (corresponding to the correct Max-Cut solution) and do not change further.

From Fig. ~\ref{TTS1}(a), (b), we observe that TTS obtained for spin oscillators and generalized PBOs follow the same trend, i.e., while there are some peaks, TTS overall doesn't increase with graph size. The peaks (long TTS) correspond to oscillators going through meta-stable phases, corresponding to local minima in the energy landscape (sub-optimal partitioning of the graph essentially), and then reaching the correct solution (corresponding to global energy minimum). Such phase evolution for spin oscillators (obtained through solving Slavin's model) has been shown in Fig. ~\ref{TTS2} for the 10-node random cubic graph, for which a TTS peak is observed in Fig. ~\ref{TTS1}(b) for the spin oscillator case.  The meta-stable states and sub-optimal partitionings of the graph corresponding to them are also shown in Fig. ~\ref{TTS2}. In contrast to that, short TTS (which is the case for most graph instances considered) corresponds to oscillators directly going to the final steady state which yields correct solution, without going through meta-stable states. Such phase evolution of spin oscillators has been shown for 18-node random cubic and 20-node Mobius Ladder cases in Section 6 of Supplementary Information.


To calculate TTS for the GW algorithm, we measure the time taken for solving the SDP relaxation and one random hyperplane measurement on a conventional digital computer with the following specifications: $11^{\mathrm{th}}$ Gen Intel(R) Core(TM) $\mathrm{i}5$-$1135$ processor and $16$ $\mathrm{GB}$ memory. We measure 100 such times by repeating the above process for a given graph instance and report the average time (and standard deviation) in milliseconds in Fig. \ref{TTS3}(b). 

For the four kinds of graphs considered here, TTS is plotted as a function of graph size for the Kuramoto model of generalized PBOs and classical GW algorithm in Fig. ~\ref{TTS3}. We have not included QAOA in this comparison because the correctness of the solution for QAOA is way below satisfactory when compared to the other methods, as discussed above. Fig. ~\ref{TTS3}(b) shows that TTS for GW grows polynomially with graph size for all kinds of graphs, which agrees with the report by Haribara \textit{et al} \cite{GWvsSAvsCIM}. But the TTS of PBOs doesn't grow overall with graph size (Fig. ~\ref{TTS3}(a)), despite a few peaks that have been explained earlier. This trend in TTS can be a major advantage of TTS compared to GW.

That TTS doesn't grow with graph size has been observed for optical coherent Ising machines (CIMs) as well in the report by Haribara \textit{et al} \cite{GWvsSAvsCIM}. But CIMs have major disadvantages like the use of a long fibre ring cavity and a power-hungry field-programmable gate array (FPGA) to implement coupling, which is not the case with spintronic oscillators \cite{MIT2}.

\section{Conclusion}

Thus, our results in the paper show that spintronic oscillators are much superior to gate-based NISQ-era quantum hardware (which implements QAOA) \cite{NISQReview} for solving combinatorial optimization problems like Max-Cut. Not only it's true that spin oscillators work at room temperature while quantum computers need milli-Kelvins to operate \cite{NISQReview}  but also spin oscillators offer much higher solution accuracy (AR and SP) than QAOA. 


The QAOA results here, against which we have compared oscillator results, are obtained from quantum circuit simulations, without incorporating any kind of quantum decoherence/ noise. It has been shown that QAOA performance goes down drastically when QAOA is implemented experimentally, or experimentally bench-marked noise is incorporated in QAOA simulations \cite{QAOA_Noise,QAOA_ICQCE1,QAOA_ICQCE2}. In comparison to that, oscillator-based computing has been found to be much more immune to noise. In fact, it has been shown that noise can actually improve AR and SP for PBOs because noise helps the system get out of local minima and reach the global minimum \cite{Jaijeet1,Jaijeet2,Csaba_Noise}. Noise immunity is, hence, an added advantage of oscillators over QAOA.

Our results further show that PBOs (including spin oscillators) may offer better TTS compared to classical GW algorithm, run on conventional computers and routinely used for solving combinatorial optimization problems currently. TTS for GW grows polynomially with graph size, while TTS for spin oscillators, and PBOs in general (modeled through the Kuramoto model), doesn't for the graph instances we have explored in this work.  
\newpage

\renewcommand{\thefigure}{S\arabic{figure}}
\setcounter{figure}{0}  

\begin{center} 
{\huge\textbf{Supplementary Information for \\ `Phase-Binarized Spintronic Oscillators for \\ Combinatorial Optimization, and \\ Comparison with Alternative Classical \\ and Quantum Methods'}}
\end{center}

\section*{Section 1: Dipole-Coupled Spin Hall Nano-Oscillators (SHNOs) in a Square Configuration}

\begin{figure*}[h]
\centering
\includegraphics[width=0.99\textwidth]{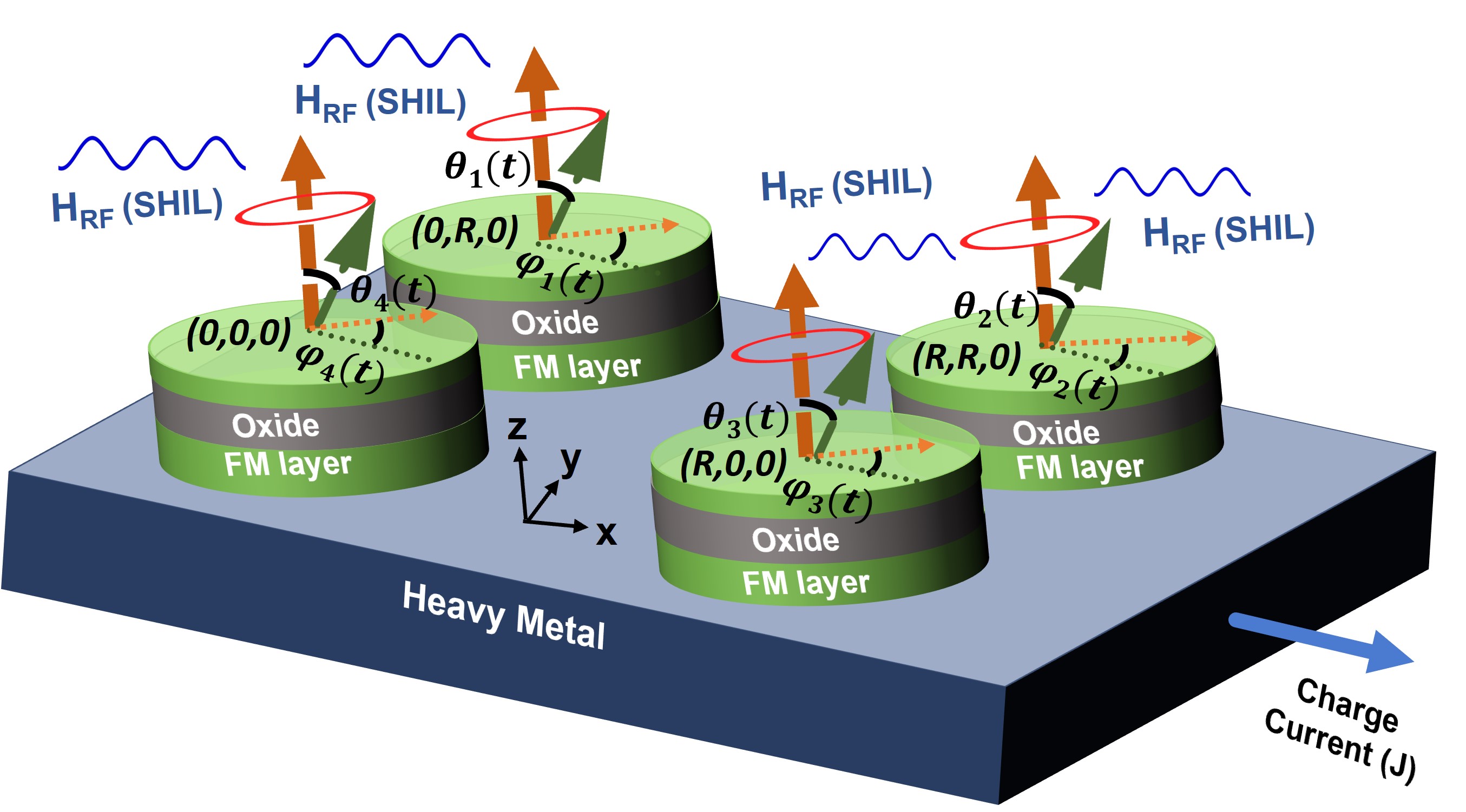}
\caption{An array of dipole-coupled heavy metal/ ferromagnetic metal (FM)/ oxide heterostructure-based SHNOs with SHNOs placed at the corners of a square. The dynamics of each SHNO is given by the time evolution of its macro-spin vector, represented by $\theta(t)$ and $\phi(t)$ in spherical coordinates as shown. Uniform RF magnetic field $H_{RF}$ is applied on the SHNOs to induce sub-harmonic injection locking (SHIL) .} 
\label{SHNOSquareSchematic}
\end{figure*}

First, we consider an array of four SHNOs where the SHNOs are placed at the corners of a square of edge length $R$, as shown in Fig. ~\ref{SHNOSquareSchematic}. Let SHNO1 be placed at position (0,$R$,0), SHNO2 be at ($R$,$R$,0), SHNO3 be at position ($R$,0,0), and SHNO4 be at (0,0,0), where $R$ is the distance between any two adjacent SHNOs which is taken to be 225 nm in our simulations.

Effective magnetic field experienced by SHNOs is given by $\vec{H}_{eff}$ ($= H_{eff}^{x}\hat{x}+H_{eff}^{y}\hat{y}+H_{eff}^{z}\hat{z}$). For a single SHNO that exhibits perpendicular magnetic anisotropy (PMA) (corresponding field: $H_k$, and where a constant applied field $H_{DC}$ is applied in the out-of-plane direction (z),  the effective magnetic field in different directions is given as follows: $H_{eff}^{x}(t)=0$, $H_{eff}^{y}(t)=0$, and   $H_{eff}^{z}(t)=H_{DC}+(H_k-\mu_{0}M_{sat})cos(\theta(t))$ \cite{Taniguchi}. ($M_{sat}$: saturation magnetization.) Here, $\theta(t)$ is the angle that the macro-spin vector of the SHNO makes with the out-of-plane $z$ axis (polar angle), as mentioned in the main text.  

To obtain phase binarization through sub-harmonic injection locking (SHIL), an extra RF external field $H_{RF}$ of strength $B_0$ (5 mT in our simulation) is applied vertically (along $z$ axis) to all SHNOs. The frequency of this external RF field ($F_e$) is kept double the natural frequency of oscillators, as mentioned in the main text.
The dipolar field is 0 in the z-direction for all SHNOs so the effective field in this direction will be same for all SHNOs. 

So, for the case of SHIL, the effective field in the out-of-plane $z$ direction is given by:

$H_{eff}^{n,z}(t)=H_{appl}+(H_k-\mu_{0}M_{sat})cos(\theta_n(t))+B_{0}sin(2\pi F_{e} t)$.

The effective field in the x and y directions due to the dipole field of other SHNOs can be given by the following equations. Expressions for the dipole field from the report by \cite{Amin} \textit{et al} have been used here. The complete derivation is provided in the reports by Garg \textit{et al} \cite{Spin_DB1} and Hemadri Bhotla \textit{et al} \cite{Spin_DB2}.

\begin{multline}
H_{eff}^{1,x}(t) =  K_1 (2 sin(\theta_2(t)) cos(\phi_2(t)) -  sin(\theta_4(t)) cos(\phi_4(t))) + \\
              K_2 sin(\theta_3(t)) (\sqrt{2} cos(\phi_3(t) +  \pi/4) 
              - 1/\sqrt{2} sin(\phi_3(t) + \pi/4))
\label{RFfield_1x}    
\end{multline}

\begin{multline}
H_{eff}^{1,y}(t) = 
K_1 (- sin(\theta_2(t)) sin(\phi_2(t)) + 2 sin(\theta_4(t)) sin(\phi_4(t))) - \\
              K_2 sin(\theta_3(t)) (\sqrt{2} cos(\phi_3(t) + \pi/4) + 1/\sqrt{2} sin(\phi_3(t) + \pi/4))
\label{RFfield_1y}    
\end{multline}

\begin{multline}
H_{eff}^{2,x}(t) =  K_1 (2 sin(\theta_1(t)) cos(\phi_1(t)) -  sin(\theta_3(t)) cos(\phi_3(t))) + \\
              K_2 sin(\theta_4(t)) (\sqrt{2} cos(-\phi_4(t) +  \pi/4) 
              - 1/\sqrt{2} sin(-\phi_4(t) + \pi/4))
\label{RFfield_2x}    
\end{multline}

\begin{multline}
H_{eff}^{2,y}(t) = 
K_1 (- sin(\theta_1(t)) sin(\phi_1(t)) + 2 sin(\theta_3(t)) sin(\phi_3(t))) + \\
              K_2 sin(\theta_4(t)) (\sqrt{2} cos(-\phi_4(t) + \pi/4) + 1/\sqrt{2} sin(-\phi_4(t) + \pi/4))
\label{RFfield_2y}    
\end{multline}

\begin{multline}
H_{eff}^{3,x}(t) =  K_1 (2 sin(\theta_4(t)) cos(\phi_4(t)) -  sin(\theta_2(t)) cos(\phi_2(t))) + \\
              K_2 sin(\theta_1(t)) (\sqrt{2} cos(\phi_1(t) +  \pi/4) 
              - 1/\sqrt{2} sin(\phi_1(t) + \pi/4))
\label{RFfield_3x}    
\end{multline}

\begin{multline}
H_{eff}^{3,y}(t) = 
K_1 (- sin(\theta_4(t)) sin(\phi_4(t)) + 2 sin(\theta_2(t)) sin(\phi_2(t))) - \\
              K_2 sin(\theta_1(t)) (\sqrt{2} cos(-\phi_1(t) + \pi/4) + 1/\sqrt{2} sin(\phi_1(t) + \pi/4))
\label{RFfield_3y}    
\end{multline}

\begin{multline}
H_{eff}^{4,x}(t) =  K_1 (2 sin(\theta_3(t)) cos(\phi_3(t)) -  sin(\theta_1(t)) cos(\phi_1(t))) + \\
              K_2 sin(\theta_2(t)) (\sqrt{2} cos(-\phi_2(t) +  \pi/4) 
              - 1/\sqrt{2} sin(-\phi_2(t) + \pi/4))
\label{RFfield_4x}    
\end{multline}

\begin{multline}
H_{eff}^{4,y}(t) = 
K_1 (- sin(\theta_3(t)) sin(\phi_3(t)) + 2 sin(\theta_1(t)) sin(\phi_1(t))) + \\
              K_2 sin(\theta_2(t)) (\sqrt{2} cos(-\phi_2(t) + \pi/4) + 1/\sqrt{2} sin(-\phi_2(t) + \pi/4))
\label{RFfield_4y}    
\end{multline}

Here $K_1 = \frac{\mu_{0}M_{sat}V} {4\pi(R^3)}$ and $K_2 = K_1/ 2 \sqrt{2}$ ($V$: volume of the ferromagnetic layer of the SHNO).

The above expressions for dipole fields and net effective fields experienced by the SHNOs are used in our numerical LLGS model of SHNOs in a square array, as mentioned in the main text.

\section*{Section 2: Four Dipole-Coupled SHNOs in a Line/ Chain Configuration}

\begin{figure*}
\centering
\includegraphics[width=0.99\textwidth]{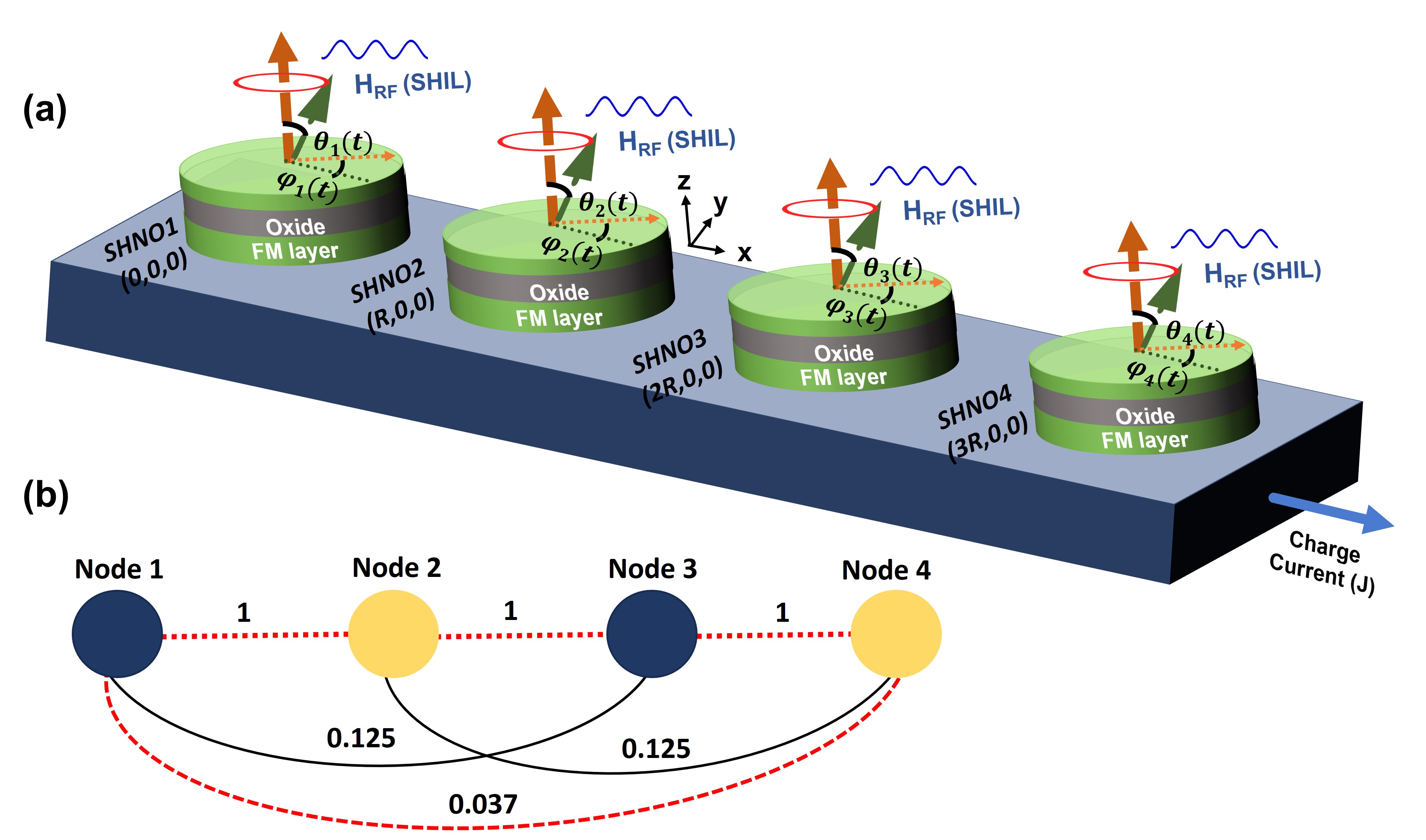}
\caption{An array of dipole-coupled heavy metal/ ferromagnetic metal (FM)/ oxide heterostructure-based SHNOs with SHNOs arranged in a line/ chain. The dynamics of each SHNO is given by the time evolution of its macro-spin vector, represented by $\theta(t)$ and $\phi(t)$ in spherical coordinates as shown. Uniform RF magnetic field $H_{RF}$ is applied on the SHNOs to induce sub-harmonic injection locking (SHIL) (b) The weighted complete graph corresponding to the array in (a), with each edge weight value chosen in proportion to the magnitude of dipole field acting between the two oscillators in (a) corresponding to the two nodes of the edge. Partitioning corresponding to the Max-Cut solution is also shown: nodes 1 and 3 belong to the first partition (blue circles), nodes 2 and 4 belong to the second partition (yellow circles), and edges connecting nodes of opposite partitions which contribute to the Max-Cut score (3.037), are highlighted by red dashed lines.} 
\label{SHNOLineSchematic}
\end{figure*}

We consider in this section an array of four SHNOs where the SHNOs are placed in a line, as shown in Fig. ~\ref{SHNOLineSchematic} (a). Let SHNO1 be placed at position (0,0,0), SHNO2 be at ($R$,0,0), SHNO3 be at position ($2R$,0,0), and SHNO4 be at ($3R$,0,0), where $R$ is the distance between any two adjacent SHNOs which is kept to be same as earlier. The effective field in z-direction remains the same as earlier but it changes in the x and y directions due to changes in dipole fields. These are given by -

\vspace{-1.5em}
\begin{multline}
H_{eff}^{1,x}(t) =  k (2 cos(\phi_2(t)) sin(\theta_2(t)) \\
            +  k_1 (2 cos(\phi_3(t)) sin(\theta_3(t)) + k_2 (2 cos(\phi_4(t)) sin(\theta_4(t)) 
\label{RFfield_Row_1x}    
\end{multline}

\vspace{-1.5em}

\begin{multline}
H_{eff}^{1,y}(t) =  - k ( sin(\phi_2(t)) sin(\theta_2(t)) \\
            -  k_1 (sin(\phi_3(t)) sin(\theta_3(t)) - k_2 (sin(\phi_4(t)) sin(\theta_4(t)) 
\label{RFfield_Row_1y}    
\end{multline}

\begin{multline}
H_{eff}^{2,x}(t) =  k (2 cos(\phi_1(t)) sin(\theta_1(t)) \\
            +  k (2 cos(\phi_3(t)) sin(\theta_3(t)) + k_1 (2 cos(\phi_4(t)) sin(\theta_4(t)) 
\label{RFfield_Row_2x}    
\end{multline}

\vspace{-1.5em}

\begin{multline}
H_{eff}^{2,y}(t) =  - k ( sin(\phi_1(t)) sin(\theta_1(t)) \\
            -  k (sin(\phi_3(t)) sin(\theta_3(t)) - k_1 (sin(\phi_4(t)) sin(\theta_4(t)) 
\label{RFfield_Row_2y}    
\end{multline}

\vspace{-1.5em}

\begin{multline}
H_{eff}^{3,x}(t) =  k (2 cos(\phi_2(t)) sin(\theta_2(t)) \\
            +  k (2 cos(\phi_4(t)) sin(\theta_4(t)) + k_1 (2 cos(\phi_1(t)) sin(\theta_1(t)) 
\label{RFfield_Row_3x}    
\end{multline}

\vspace{-1.5em}

\begin{multline}
H_{eff}^{3,y}(t) =  - k ( sin(\phi_2(t)) sin(\theta_2(t)) \\
            -  k (sin(\phi_4(t)) sin(\theta_4(t)) - k_1 (sin(\phi_1(t)) sin(\theta_1(t)) 
\label{RFfield_Row_3y}    
\end{multline}

\vspace{-1.5em}

\begin{multline}
H_{eff}^{4,x}(t) =  k (2 cos(\phi_3(t)) sin(\theta_3(t)) \\
            +  k_1 (2 cos(\phi_2(t)) sin(\theta_2(t)) + k_2 (2 cos(\phi_1(t)) sin(\theta_1(t)) 
\label{RFfield_Row_4x}    
\end{multline}

\vspace{-1.5em}

\begin{multline}
H_{eff}^{4,y}(t) =  - k ( sin(\phi_3(t)) sin(\theta_3(t)) \\
            -  k_1 (sin(\phi_2(t)) sin(\theta_2(t)) - k_2 (sin(\phi_1(t)) sin(\theta_1(t)) 
\label{RFfield_Row_4y}    
\end{multline}

Here $k = \frac{\mu_{0}M_{sat}V} {4\pi(R^3)}$, $k_1 = k/8$ and $k_2 = k/27$ 

\begin{figure*}
\centering
\includegraphics[width=0.99\textwidth]{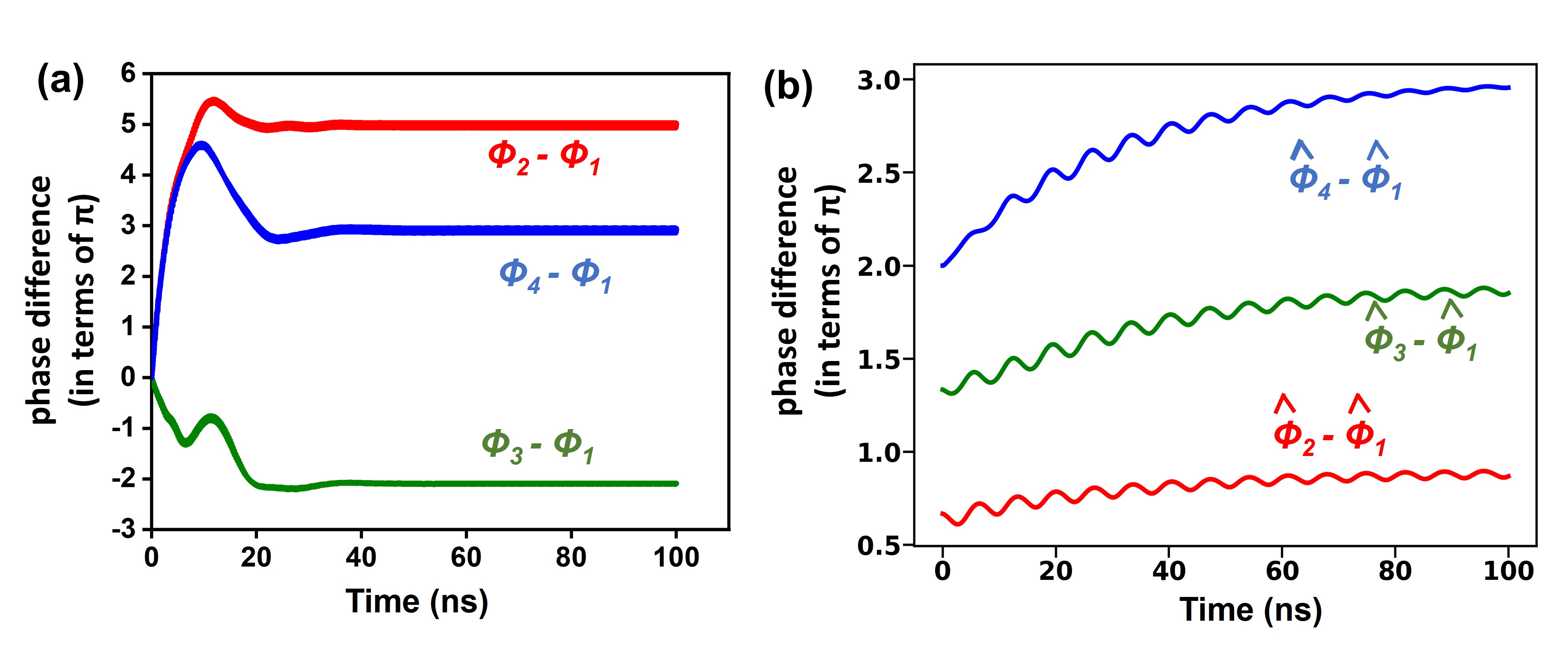}
\caption{With respect to the oscillator array of Fig. ~\ref{SHNOLineSchematic}(a), the phase difference between oscillator 2 and oscillator 1 (red plot), oscillator 3 and oscillator 1 (green plot), and oscillator 4 and oscillator 1 (blue plot) plotted here are functions of time as obtained from: (a) LLGS model of the spin oscillator, (b) Slavin's model of the spin oscillator. Partitioning of the graph depends upon whether phase differences are odd or even multiples of $\pi$. As can be seen from (a) and (b), both LLGS and Slavin's model yield correct graph partitioning and Max-Cut score (3.037), as shown in Fig. 2(a).} 
\label{SHNOLinePhases}
\end{figure*}

The above-written expressions of dipole fields and net effective fields are used in the main equations (as before) to study the phase dynamics of oscillators arranged in a row. From Fig. ~\ref{SHNOLinePhases} (a) and (b), this can be verified that both single domain and Slavin's model give the same result as given by classical brute solver (Fig. ~\ref{SHNOLineSchematic} (b)) irrespective of different geometries.  

\section*{Section 3: Descriptions of the Four Types of Graphs Used Here}

\begin{figure}[!t]
     \centering     \includegraphics[width=0.99\textwidth]{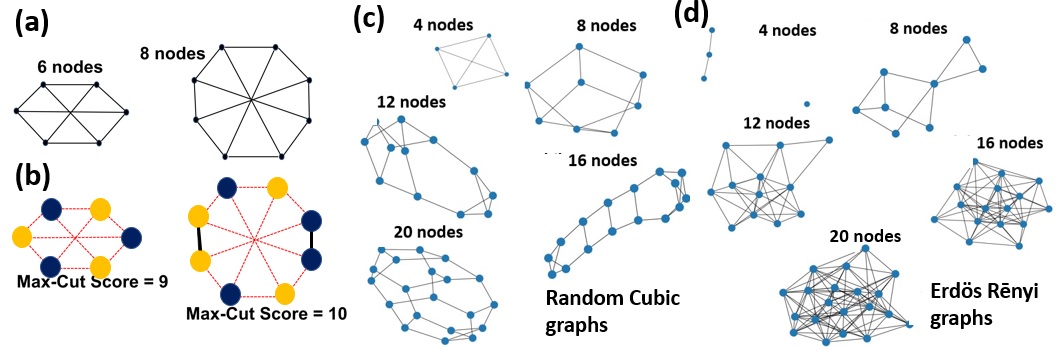}
    \caption{(a) 6-node Mobius ladder and 8-node Mobius ladder graphs (unweighted) are shown without showing any partitioning. (b) Partitioning of the graphs in (a) such that the corresponding score for the Max-Cut problem is maximized is shown here. Nodes belonging to one partition are shown as dark blue circles, nodes belonging to the other partition are shown as yellow circles, and the edges that contribute to the final score that is to be maximized are shown as red dashed lines. The corresponding score is also mentioned. (c) Various instances of the random cubic graph used in this work (d) Various instances of the Erd{\"o}s R{\'e}nyi graph used in this work 
         }
    \label{GraphInstances}
\end{figure}

\textbf{Mobius ladder graph:} A Mobius Ladder graph (shown in Fig. ~\ref{GraphInstances}(a)) of graph size $N$ ($N$ is a positive even integer) is an undirected unweighted 3-regular (also known as cubic) graph with $N$ node, for which every node $i$ has edges with only three other nodes \cite{Ref_MobiusLadder}:
$$
\begin{cases}
i-1, i+1, (\frac{N}{2})+i, \hspace{1em} \text{if} \hspace{1em}  2 \leq i \leq \frac{N}{2} \\
i-1, i+1, i-(\frac{N}{2}), \hspace{1em} \text{if} \hspace{1em}  (\frac{N}{2})+1 \leq i < N \\
n, i+1, (\frac{N}{2})+i \hspace{1em} \text{if} \hspace{1em} i = 1\\
i-1, 1, i-(\frac{N}{2}) \hspace{1em} \text{if} \hspace{1em} i = N  \\
\end{cases}
$$

With all other nodes, node $i$ doesn't have edges. Fig. ~\ref{GraphInstances}(a) shows Mobius ladder graphs of graph size (number of nodes) 6 and 8. Correct partitioning of these graphs, yielding the correct Max-Cut score, is shown in Fig. ~\ref{GraphInstances}(b). In our paper, we work on Mobius ladder graphs with graph size $N$ (even integer) from 4 to 20.

\textbf{Random cubic graph:}  Any random cubic graph we use here is an undirected unweighted 3-regular graph where each node is randomly connected to three other nodes \cite{Ref_RandomCubic}. In our paper, for each even value of graph size $N$ from 4 to 20, we generate a random cubic graph using the NetworkX Python package \cite{SciPyProceedings_11,Ref_RandomCubic,gen_random_reg}. The specific 4-node, 8-node, 12-node, 16-node, and 20-node random cubic graph instances we generate for this work are shown in Fig. ~\ref{GraphInstances}(c).

\textbf{Erd{\"o}s R{\'e}nyi graph:} Any Erd{\"o}s R{\'e}nyi graph we use here is given by $G(N,p)$ and is an undirected unweighted graph of $N$ nodes, where the probability of any edge between two nodes to exist (also known as edge probability) is independent of any other edge to exist and is given by $p$ ($0 \leq p \leq 1$) \cite{Ref_ErdosRenyi}. Thus, the expected number of nodes for this graph is $p$ ${N}\choose{k}$. In our paper, for each value of $N$ from 4 to 20, we generate an Erd{\"o}s R{\'e}nyi graph given by $G(N,0.5)$.  We choose edge probability $p$= 0.5 because it corresponds to Erd{\"o}s R{\'e}nyi graphs of the highest difficulty level in terms of solving the Max-Cut problem. Graphs with both very low level of connectivity (low value of $p$) and very high level of connectivity (high value of $p$) have been found to be much easier Max-Cut problem instances compared to that \cite{QAOA_ErdosRenyi}. The specific 4-node, 8-node, 12-node, 16-node, and 20-node Erd{\"o}s R{\'e}nyi graph instances we generate for this work are shown in Fig. ~\ref{GraphInstances}(d).

\textbf{Complete graph:} Any complete graph we use here is an undirected but weighted graph, where every node is connected to every other node by an edge \cite{Ref_CompleteGraph,QAOA_CompleteGraph}. A weight is assigned randomly to every edge, independent of other edges, with each weight sampled uniformly at random from [0,1].

\section*{Section 4: Random Hyperplane Cuts for the Classical GW Algorithm}

\begin{figure*}
\centering
\includegraphics[width=0.99\textwidth]{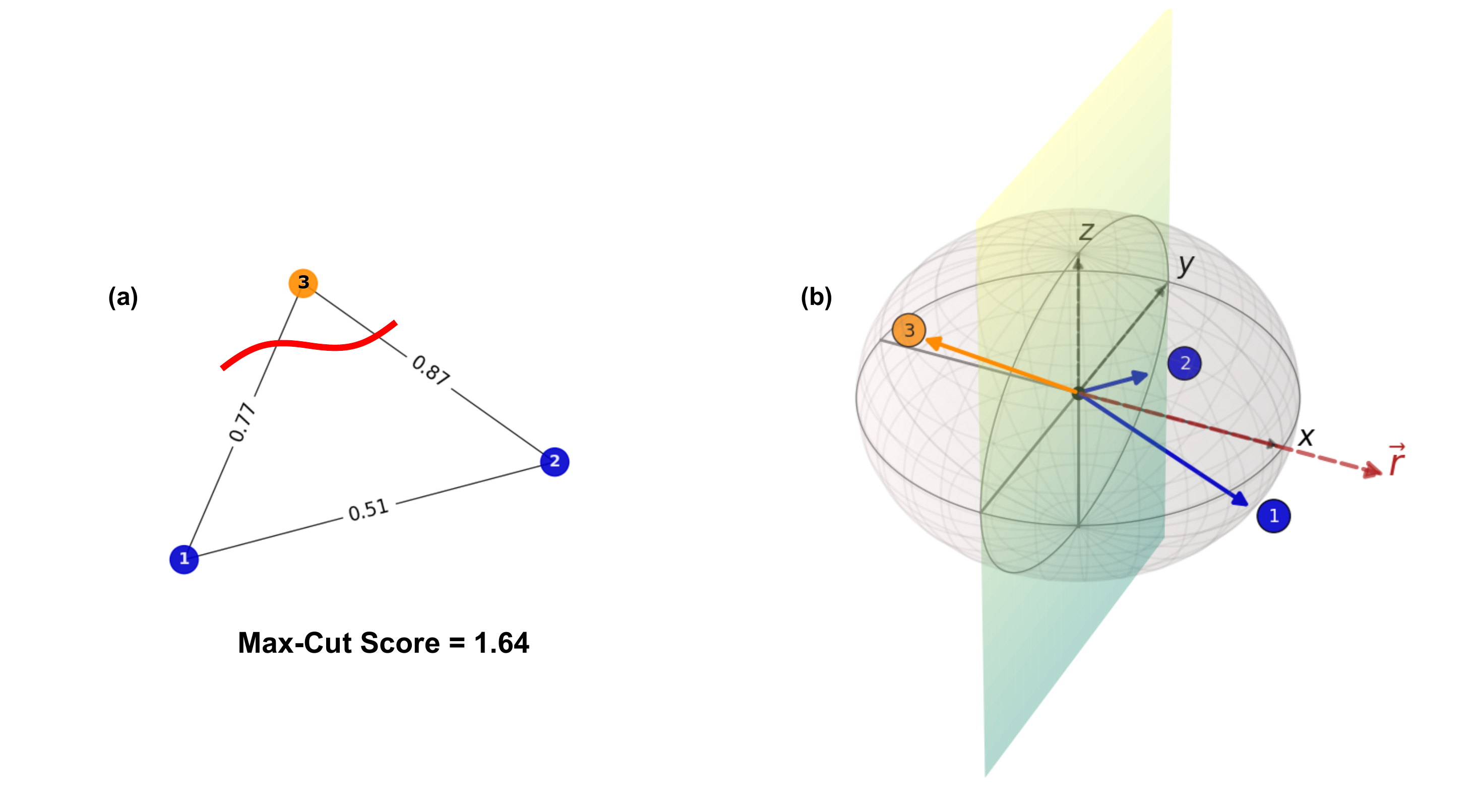}
\caption{(a) 3-node weighted complete graph is shown with a MaxCut score of $1.64$. The nodes $1$ and $2$ are in one partition, while node $3$ belongs in the other partition. (b) GW's measurement step is depicted here. Solving the SDP formulation of the graph in (a) outputs three optimal vectors $\Vec{x}_{\text{opt}}$, one for each vertex. One of the random hyperplanes is given by the $y$-$z$ plane with its normal vector $\Vec{r}$. Since the vectors $\Vec{x}_{\text{opt}}$ for nodes $1$ and $2$ have a positive inner product with $\Vec{r}$, they are mapped to $+1$, while the node $3$ is assigned $-1$. The mean approximation ratio from $1024$ random measurements is $0.920\pm0.095$. } 
\label{GW_hyperplane_supple}
\end{figure*}

A random hyperplane cut for a 3-node complete weighted graph, using GW algorithm discussed in the main text, is shown in Fig. ~\ref{GW_hyperplane_supple}.

\section*{Section 5: QAOA Implementation Details}

\begin{figure*}
     \centering
\includegraphics[width=0.99\textwidth]{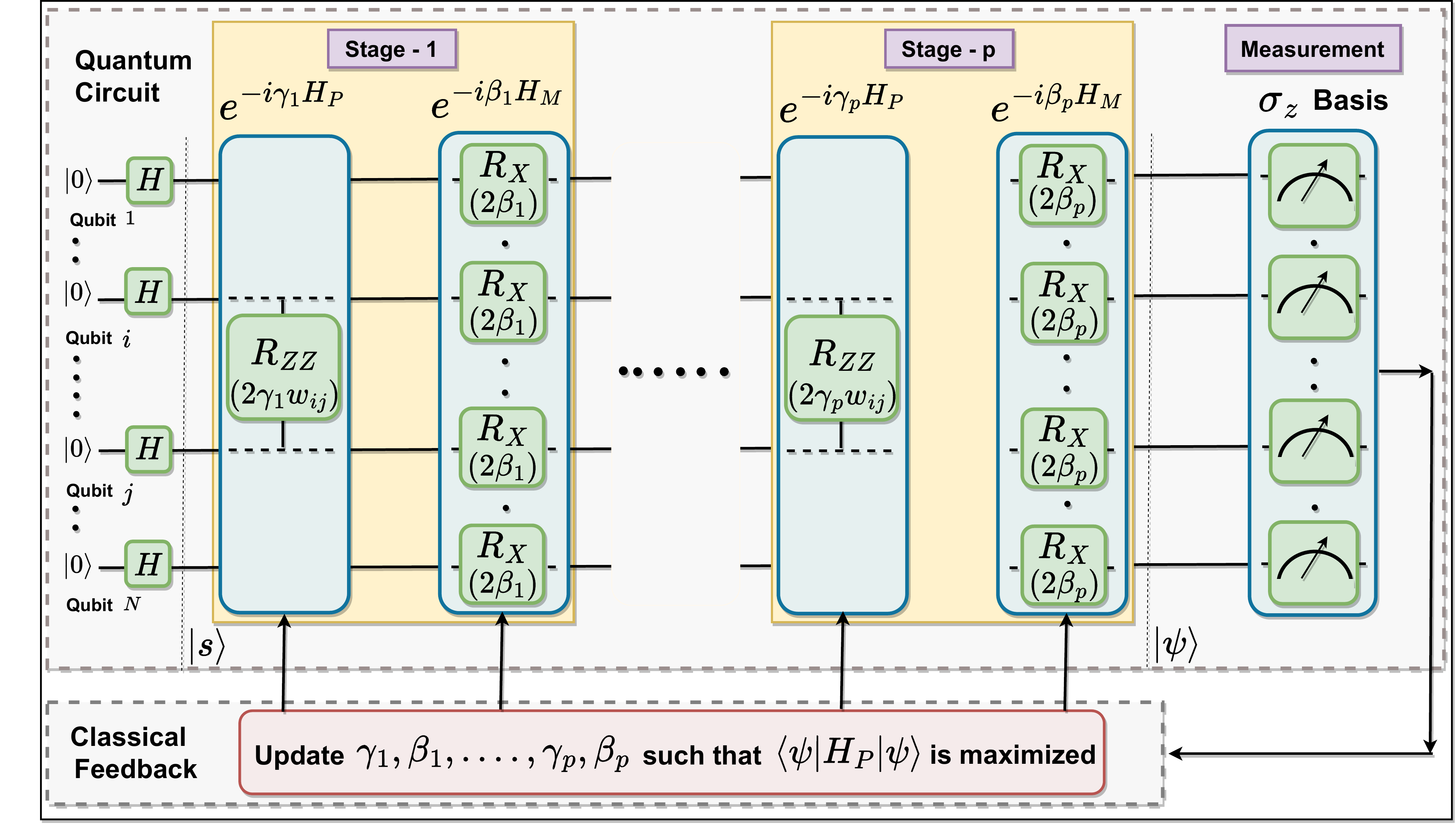}
\caption{Schematic of Quantum Approximation Optimization Algorithm (QAOA), depicting the quantum circuit with $p$ stages for the forward pass accompanied by a classical feedback loop designed to update the parameters of the quantum circuit: ($\beta_1$, $\gamma_1$, ... $\beta_p$, $\gamma_p$).
Total number of qubits $N$ is identical to the number of nodes for a given graph instance. All symbols in the figure and circuit details are explained in the text. Here, the single-qubit gate $H$ corresponds to the Hadamard gate. The unitary operator $e^{-i\beta H_M}$ can be decomposed into single-qubit $R_X(2\beta)$ gates applied on each qubit. For an unweighted graph, the unitary operator $e^{-i\gamma H_P}$ can be decomposed into two-qubit $R_{ZZ}(2\gamma)$ gates applied on every two qubits $i$ and $j$ if there exists a graph edge connecting nodes $i$ and $j$, and no such gate on qubits $i$ and $j$ if there is no edge between $i$ and $j$. However, for a weighted graph, $e^{-i\gamma H_P}$ can be decomposed into two-qubit $R_{ZZ}(2\gamma w_{i,j})$ gates applied on every two qubits $i$ and $j$, where $w_{i,j}$ is the weight of an edge connecting graph nodes $i$ and $j$.
}
\label{QAOA_Circuit}
\end{figure*}

The quantum circuit designed and used to implement QAOA for this work is shown in Fig. ~\ref{QAOA_Circuit}. We have used Python 3 with IBM's Qiskit quantum simulation package \cite{Qiskit} to implement QAOA. The forward computation, as given by equations in the main text and as shown in Fig. ~\ref{QAOA_Circuit}, is simulated using the Aer Simulator within the Qiskit package, without any noise model added. We have used 10 stages only in our QAOA circuit of Fig. ~\ref{QAOA_Circuit} for all the graph instances we use ($p$=10), which is a reasonable number (very few stages leads to low performance, having too many stages makes the Qiskit simulation slow and corresponding hardware implementation very noisy). Nelder-Mead optimization algorithm \cite{NelderMead}, provided in the SciPy package, has been used as the classical optimizer to optimize the expectation value ($\bra{\psi}H_P\ket{\psi}$). ($\beta_1$, $\gamma_1$, $\beta_2$, $\gamma_2$, ... $\beta_p$, $\gamma_p$) are updated over several iterations as long as the difference in expectation value between two consecutive iterations is greater than $10^{-4}$. We choose Nelder-Mead method over other optimization methods because it has been shown previously in several reports that Nelder-Mead method \cite{NelderMead} is a very effective optimization method for these kind of problems \cite{QAOA_ErdosRenyi,QuantumAnnealingvsQAOA1,QAOA_NelderMead1,QAOA_NelderMead2}.

Before the first iteration, initial values of ($\beta_1$, $\beta_2$, ... $\beta_p$) are sampled uniformly at random from [0,$\pi$] because $e^{-i\beta H_M}$ is a periodic function of $\beta$ with a periodicity of $\pi$. Initial values of ($\gamma_1$, $\gamma_2$, ... $\gamma_p$) are sampled uniformly at random from [0,$2\pi$] because $e^{-i\gamma H_P}$ is a periodic function of $\gamma$ with a periodicity of $2\pi$ \cite{QAOA_Farhi,QAOA_ErdosRenyi,QAOA_CompleteGraph,QAOA_NelderMead1,QAOA_NelderMead2}. 

Once the difference between the expectation value ($\bra{\psi}H_P\ket{\psi}$)is less than $10^{-4}$, the iterative parameter update process ends and we obtain the final values of ($\beta_1$, $\gamma_1$, $\beta_2$, $\gamma_2$, ... $\beta_p$, $\gamma_p$).

\section*{Section 6: Phase Evolution of Spin Oscillators}

\begin{figure*}
    \centering
\includegraphics[width=0.99\textwidth]{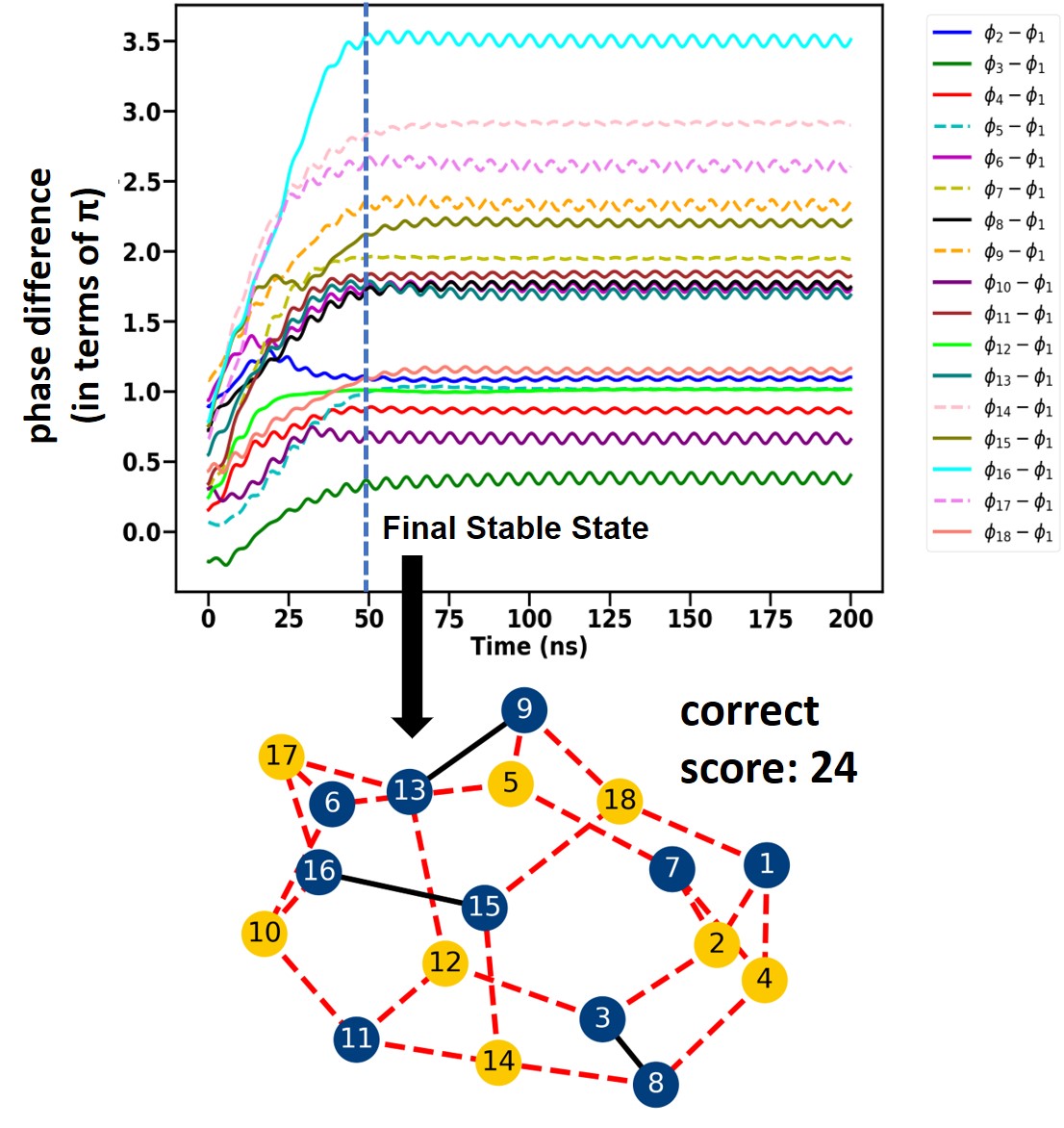}
\caption{Evolution of the phases of spin oscillators, modelled by Slavin's model, over time corresponding to the 18-node random cubic graph instance we have used in this work. No meta-stable state is formed in this case. Partitioning for the final stable state (correct solution) is shown.
\vspace{-1em}}
\label{TTS_18nodeRC}
\end{figure*}

\begin{figure*}
    \centering
\includegraphics[width=0.99\textwidth]{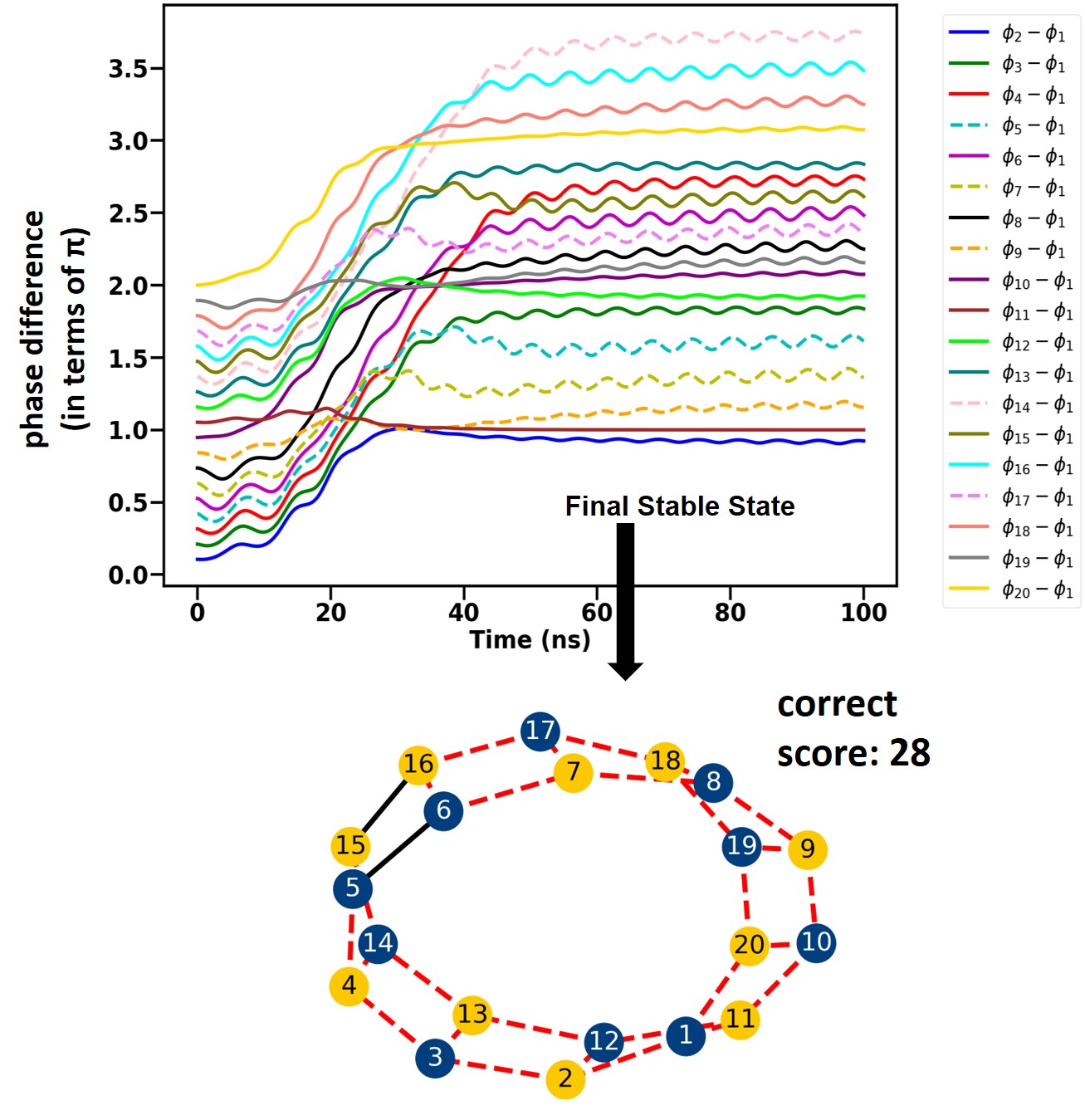}
\caption{Evolution of the phases of spin oscillators, modelled by Slavin's model, over time corresponding to the 20-node Mobius Ladder graph instance we have used in this work. No meta-stable state is formed in this case. Partitioning for the final stable state (correct solution) is shown.
\vspace{-1em}}
\label{TTS_20nodeML}
\end{figure*}

In Fig. ~\ref{TTS_18nodeRC} and ~\ref{TTS_20nodeML}, the phase evolution of spin oscillators over time, as obtained from Slavin's model, is shown for two cases: 18-node random cubic graph instance we have considered here and 20-node Mobius Ladder graph. In both these cases, TTS is short (no peak observed in Fig. 5 of the main text). From Fig. ~\ref{TTS_18nodeRC} and ~\ref{TTS_20nodeML}, we observe that in these cases, as expected, the oscillators do not go through any meta-stable state (unlike the case of 10-node random cubic graph in Fig. 6 of the main paper). The oscillators straight away evolve into their final stable state, which yields the correct Max-Cut partitioning and score.

\bibliographystyle{ieeetr}
\bibliography{arxiv}

\providecommand{\noopsort}[1]{}\providecommand{\singleletter}[1]{#1}%
\begin{thebibliography}{10}

\bibitem{CombApp1}
V.~Chvatal, {\em Combinatorial Optimization: Methods and Applications}.
\newblock IOS Press, 2011.

\bibitem{CombApp2}
R.~Unger and J.~Moult, ``Finding the lowest free energy conformation of a
  protein is a np-hard problem: Proof and implications,'' {\em Bulletin of
  Mathematical Biology}, vol.~55, pp.~1183--1195, 1993.

\bibitem{RingOscApp}
I.~Ahmed, P.~Chiu, W.~Moy, and C.~Kim, ``A probabilistic compute fabric based
  on coupled ring oscillators for solving combinatorial optimization
  problems,'' {\em IEEE Journal of Solid-State Circuits}, vol.~56,
  pp.~2870--2880, 2021.

\bibitem{GW1}
M.~X. Goemans and D.~P. Williamson, ``Improved approximation algorithms for
  maximum cut and satisfiability problems using semidefinite programming,''
  {\em Journal of the Association for Computing Machinery}, vol.~42,
  pp.~1115--1145, 1995.

\bibitem{GWvsSAvsCIM}
S.~U. Y.~Haribara and Y.~Yamamoto, ``A coherent ising machine for max-cut
  problems: Performance evaluation against semidefinite programming and
  simulated annealing,'' {\em In: Yamamoto, Y., Semba, K. (eds) Principles and
  Methods of Quantum Information Technologies, Lecture Notes in Physics
  (Springer, Tokyo)}, vol.~911, 2016.

\bibitem{NPCompleteIsingFormulation}
A.~Lucas, ``Ising formulations of many np problems,'' {\em Front. Phys.},
  vol.~2, p.~5, 2014.

\bibitem{IsingComputingReview}
N.~Mohseni, P.~L. McMahon, and T.~Byrnes, ``Ising machines as hardware solvers
  of combinatorial optimization problems,'' {\em Nat. Rev. Phys.}, vol.~4,
  pp.~363--379, 2022.

\bibitem{Spin_Akerman1}
A.~Houshang, M.~Zahedinejad, S.~Muralidhar, J.~Chechinski, R.~Khymyn,
  M.~Rajabali, H.~Fulara, A.~A. Awad, M.~Dvornik, and J.~Akerman,
  ``Phase-binarized spin hall nano-oscillator arrays: Towards spin hall ising
  machines,'' {\em Phys. Rev. Appl.}, vol.~17, p.~014003, 2022.

\bibitem{CoherentIMvsQA}
R.~Hamerly, T.~Inagaki, P.~L. McMahon, D.~Venturelli, A.~Marandi, T.~Onodera,
  E.~Ng, C.~Langrock, K.~Inaba, T.~Honjo, K.~Enbutsu, T.~Umeki, R.~Kasahara,
  S.~Utsunomiya, S.~Kako, K.~ichi Kawarabayashi, R.~L. Byer, M.~M. Fejer,
  H.~Mabuchi, D.~Englund, E.~Rieffel, H.~Takesue, and Y.~Yamamoto,
  ``Experimental investigation of performance differences between coherent
  ising machines and a quantum annealer,'' {\em Sci. Adv.}, vol.~5,
  p.~eaau0823, 2019.

\bibitem{NISQReview}
J.~W.~Z. Lau, K.~H. Lim, H.~Shrotriya, and L.~C. Kwek, ``Nisq computing: where
  are we and where do we go?,'' {\em AAPPS Bulletin}, vol.~27, p.~32, 2022.

\bibitem{QAOA_Farhi}
E.~Farhi, J.~Goldstone, and S.~Gutmann, ``A quantum approximate optimization
  algorithm,'' {\em arXiv preprint arXiv:1411.4028}, 2014.

\bibitem{OscillatorIMReview}
Y.~Zhang, Y.~Deng, Y.~Lin, Y.~Jiang, Y.~Dong, X.~Chen, G.~Wang, D.~Shang,
  Q.~Wang, H.~Yu, and Z.~Wang, ``Oscillator-network-based ising machine,'' {\em
  micromachines}, vol.~13, p.~1016, 2022.

\bibitem{SKONN_NCE}
C.~Delacour, S.~Carapezzi1, G.~Boschetto, M.~Abernot, T.~Gil, N.~Azemard, and
  A.~Todri-Sanial, ``A mixed-signal oscillatory neural network for scalable
  analog computations in phase domain,'' {\em Neuromorph. Comput. Eng.},
  vol.~3, p.~034004, 2023.

\bibitem{Kuramoto}
Y.~Kuramoto, {\em Chemical oscillations, waves, and turbulence, vol. 19}.
\newblock Springer Berlin, Heidelberg, 2003.

\bibitem{Jaijeet1}
T.~Wang and J.~Roychowdhury, ``Oim: Oscillator-based ising machines for solving
  combinatorial optimisation problems,'' in {\em Proceedings of the
  International Conference on Unconventional Computation and Natural
  Computation}, 2019.

\bibitem{Jaijeet2}
T.~Wang, {\em Novel Computing Paradigms using Oscillators}.
\newblock {PhD} dissertation, University of California Berkeley, Department of
  Electrical Engineering and Computer Sciences, 2020.

\bibitem{Jaijeet3}
T.~Wang, L.~Wu, , and J.~Roychowdhury, ``New computational results and hardware
  prototypes for oscillator-based ising machines,'' in {\em Proceedings of the
  56th Annual Design Automation Conference (DAC) 2019}, (Las Vegas, NV, USA),
  p.~239, 2019.

\bibitem{ChrisKim1}
W.~Moy, I.~Ahmed, P.-W. Chiu, J.~Moy, S.~S. Sapatnekar, and C.~H. Kim, ``A
  1,968-node coupled ring oscillator circuit for combinatorial optimization
  problem solving,'' {\em Nat. Electron.}, vol.~5, pp.~310--317, 2023.

\bibitem{ChrisKim2}
H.~Lo, W.~Moy, H.~Yu, S.~Sapatnekar, and C.~H. Kim, ``An ising solver chip
  based on coupled ring oscillators with a 48-node all-to-all connected array
  architecture,'' {\em Nat. Electron.}, vol.~6, pp.~771--778, 2023.

\bibitem{NikhilShukla1}
A.~Mallick, M.~K. Bashar, D.~S. Truesdell, B.~H. Calhoun, S.~Joshi, and
  N.~Shukla, ``Using synchronized oscillators to compute the maximum
  independent set,'' {\em Nat. Comm.}, vol.~11, p.~4689, 2020.

\bibitem{MIT1}
N.~Shukla, A.~Parihar, E.~Freeman, H.~Paik, G.~Stone, V.~Narayanan, H.~Wen,
  Z.~Cai, V.~Gopalan, R.~Engel-Herbert, D.~G. Schlom, A.~Raychowdhury, and
  S.~Datta, ``Synchronized charge oscillations in correlated electron
  systems,'' {\em Sci. Rep.}, vol.~4, p.~4964, 2014.

\bibitem{MIT2}
S.~Dutta, A.~Khanna, A.~S. Assoa, H.~Paik, D.~Schlom, Z.~Toroczkai,
  A.~Raychowdhury, and S.~Datta, ``An ising hamiltonian solver based on coupled
  stochastic phase-transition nano-oscillators,'' {\em Nat. Electron.}, vol.~4,
  pp.~502--512, 2023.

\bibitem{Spin_Akerman2}
M.~Zahedinejad, A.~A. Awad, S.~Muralidhar, R.~Khymyn, H.~Fulara, H.~Mazraati,
  M.~Dvornik, and J.~Åkerman, ``Two-dimensional mutually synchronized spin
  hall nano-oscillator arrays for neuromorphic computing,'' {\em Nat.
  Nanotechnol.}, vol.~15, pp.~47--52, 2020.

\bibitem{Spin_Finocchio1}
G.~Finocchio, M.~D. Ventra, K.~Y. Camsari, K.~Everschor-Sitte, P.~K. Amiri, and
  Z.~Zeng, ``The promise of spintronics for unconventional computing,'' {\em J.
  Magn. Magn. Mater}, vol.~521, p.~167506, 2021.

\bibitem{Spin_Finocchio2}
A.~Grimaldi, L.~Mazza, E.~Raimondo, P.~Tullo, D.~Rodrigues, K.~Y. Camsari,
  V.~Crupi, M.~Carpentieri, V.~Puliafito, and G.~Finocchio, ``Evaluating
  spintronics-compatible implementations of ising machines,'' {\em Phys. Rev.
  Appl.}, vol.~XX, 2023.

\bibitem{Spin_DB1}
N.~Garg, S.~V.~H. Bhotla, P.~K. Muduli, and D.~Bhowmik, ``Kuramoto-model-based
  data classification using the synchronization dynamics of uniform-mode spin
  hall nano-oscillators,'' {\em Neuromorph. Comput. Eng.}, vol.~1, p.~024005,
  2021.

\bibitem{Spin_DB2}
S.~V.~H. Bhotla, N.~Garg, T.~Aggarwal, P.~K. Muduli, and D.~Bhowmik, ``An
  oscillator-synchronization-based off-line learning algorithm, with on-chip
  inference on an array of spin hall nano-oscillators,'' {\em IEEE Trans.
  Nano.}, vol.~22, pp.~136--148, 2023.

\bibitem{Spin_Ashwin}
H.~Singh, S.~Bhuktare, A.~Bose, A.~Fukushima, K.~Yakushiji, S.~Yuasa,
  H.~Kubota, and A.~A. Tulapurkar, ``Mutual synchronization of spin-torque
  nano-oscillators via oersted magnetic fields created by waveguides,'' {\em
  Phys. Rev. Appl.}, vol.~5, p.~054028, 2019.

\bibitem{Spin_Liu}
B.~C. McGoldrick, J.~Z. Sun, and L.~Liu, ``Ising machine based on electrically
  coupled spin hall nano-oscillators,'' {\em Phys. Rev. Appl.}, vol.~17,
  p.~014006, 2022.

\bibitem{Spin_Hyunsoo}
R.~Sharma, R.~Mishra, T.~Ngo, Y.~X. Guo, S.~Fukami, H.~Sato, H.~Ohno, and
  H.~Yang, ``Electrically connected spin-torque oscillators array for 2.4 ghz
  wifi band transmission and energy harvesting,'' {\em Nat. Comm.}, vol.~12,
  p.~2924, 2021.

\bibitem{Slavin}
A.~Slavin and V.~Tiberkevich, ``Nonlinear auto-oscillator theory of microwave
  generation by spin-polarized current,'' {\em IEEE Trans. Magn.}, vol.~45,
  no.~4, 2009.

\bibitem{Ref_CompleteGraph}
J.~Harris, J.~L. Hirst, and M.~Mossinghoff, {\em Combinatorics and Graph
  Theory}.
\newblock Springer, 2008.

\bibitem{Ref_MobiusLadder}
J.~P. McSorley, ``Counting structures in the möbius ladder,'' {\em Discrete
  Mathematics}, vol.~184, pp.~137--164, 1998.

\bibitem{Ref_RandomCubic}
M.~Noy, C.~Requilé, and J.~Rué, ``Further results on random cubic planar
  graphs,'' {\em Random Structures and Algorithms}, vol.~56, pp.~892--924,
  2019.

\bibitem{Ref_ErdosRenyi}
P.~Erdös and A.~Rényi., ``On random graphs i,'' {\em Publ. Math. Debrecen},
  1959.

\bibitem{Taniguchi}
T.~Taniguchi, H.~Arai, H.~Kubota, and H.~Imamura, ``Theoretical study of
  spin-torque oscillator with perpendicularly magnetized free layer,'' {\em
  IEEE Trans. Magn.}, vol.~50, no.~1, 2014.

\bibitem{Taniguchi_Expt1}
S.~Yakata, H.~Kubota, Y.~Suzuki, K.~Yakushiji, A.~Fukushima, S.~Yuasa, and
  K.~Ando, ``Influence of perpendicular magnetic anisotropy on spin-transfer
  switching current in co fe b/ mg o/ co fe b magnetic tunnel junctions,'' {\em
  J. Appl. Phys.}, vol.~105, no.~07D131, 2009.

\bibitem{Taniguchi_Expt2}
H.~Kubota, S.~Ishibashi, T.~Saruya, T.~Nozaki, A.~Fukushima, K.~Yakushiji,
  K.~Ando, Y.~Suzuki, and S.~Yuasa, ``Enhancement of perpendicular magnetic
  anisotropy in feb free layers using a thin mgo cap layer,'' {\em J. Appl.
  Phys.}, vol.~111, no.~07C723, 2012.

\bibitem{Taniguchi_Expt3}
H.~Kubota, K.~Yakushiji, A.~Fukushima, S.~Tamaru, M.~Konoto, T.~Nozaki,
  S.~Ishibashi, T.~Saruya, S.~Yuasa, T.~Taniguchi, H.~Arai, and H.~Imamura,
  ``Spin-torque oscillator based on magnetic tunnel junction with a
  perpendicularly magnetized free layer and in-plane magnetized polarizer,''
  {\em Appl. Phys. Express}, vol.~6, no.~10, 2013.

\bibitem{SHE_Pt_1}
L.~Q. Liu, T.~Moriyama, D.~C. Ralph, and R.~A. Buhrman, ``Spin torque
  ferromagnetic resonance induced by the spin hall effect,'' {\em Phys. Rev.
  Lett.}, vol.~106, no.~036601, 2011.

\bibitem{SHE_Pt_2}
L.~Q. Liu, O.~J. Lee, T.~J. Gudmundsen, D.~C. Ralph, and R.~A. Buhrman,
  ``Current-induced switching of perpendicularly magnetized magnetic layers
  using spin torque from the spin hall effect,'' {\em Phys. Rev. Lett.},
  vol.~109, no.~096602, 2012.

\bibitem{Spin_Akerman3}
D.~I. Albertsson, M.~Zahedinejad, A.~Houshang, R.~Khymyn, J.~Akerman, and
  A.~Rusu, ``Ultrafast ising machines using spin torque nano-oscillators,''
  {\em Appl. Phys. Lett.}, vol.~118, p.~112404, 2021.

\bibitem{Amin}
N.~Amin, H.~Xi, , and M.~X. Tang, ``Analysis of electromagnetic fields
  generated by a spin-torque oscillator,'' {\em IEEE Trans. Magn.}, vol.~45,
  no.~10, 2009.

\bibitem{QAOA_ErdosRenyi}
R.~Sreedhar, P.~Vikstål, M.~Svensson, A.~Ask, G.~Johansson, and L.~G.
  Álvarez, ``The quantum approximate optimization algorithm performance with
  low entanglement and high circuit depth,'' {\em arXiv:2207.03404}, 2022.

\bibitem{andersen2013cvxopt}
M.~S. Andersen, J.~Dahl, L.~Vandenberghe, {\em et~al.}, ``Cvxopt: A python
  package for convex optimization,'' {\em Available at cvxopt. org}, vol.~54,
  2013.

\bibitem{2020SciPy-NMeth}
P.~Virtanen, R.~Gommers, T.~E. Oliphant, M.~Haberland, T.~Reddy, D.~Cournapeau,
  E.~Burovski, P.~Peterson, W.~Weckesser, J.~Bright, S.~J. {van der Walt},
  M.~Brett, J.~Wilson, K.~J. Millman, N.~Mayorov, A.~R.~J. Nelson, E.~Jones,
  R.~Kern, E.~Larson, C.~J. Carey, {\.I}.~Polat, Y.~Feng, E.~W. Moore,
  J.~{VanderPlas}, D.~Laxalde, J.~Perktold, R.~Cimrman, I.~Henriksen, E.~A.
  Quintero, C.~R. Harris, A.~M. Archibald, A.~H. Ribeiro, F.~Pedregosa, P.~{van
  Mulbregt}, and {SciPy 1.0 Contributors}, ``{{SciPy} 1.0: Fundamental
  Algorithms for Scientific Computing in Python},'' {\em Nature Methods},
  vol.~17, pp.~261--272, 2020.

\bibitem{QuantumAnnealingvsQAOA1}
M.~Willsch, D.~Willsch, F.Jin, H.~D. Raedt, and K.~Michielsen, ``Benchmarking
  the quantum approximate optimization algorithm,'' {\em Quantum Inf.
  Process.}, vol.~19, p.~197, 2020.

\bibitem{Qiskit}
{Qiskit contributors}, ``Qiskit: An open-source framework for quantum
  computing,'' 2023.

\bibitem{QAOA_Tutorial}
J.~Choi and J.~Kim, ``A tutorial on quantum approximate optimization algorithm
  (qaoa): Fundamentals and applications,'' in {\em Proceedings of the 2019
  International Conference on Information and Communication Technology
  Convergence (ICTC)}, (Jeju, Korea (South)), 2019.

\bibitem{MobiusLadder_PolySoln}
K.~P. Kalinin and N.~G. Berloff, ``Computational complexity continuum within
  ising formulation of np problems,'' {\em Communications Physics}, vol.~5,
  p.~20, 2022.

\bibitem{QAOA_Noise}
M.~Harrigan, K.~Sung, M.~Neeley, and et~al, ``Quantum approximate optimization
  of non-planar graph problems on a planar superconducting processor,'' {\em
  Nature Physics}, vol.~17, pp.~332--336, 2014.

\bibitem{QAOA_ICQCE1}
R.~Shaydulin and A.~Galda, ``Error mitigation for deep quantum optimization
  circuits by leveraging problem symmetries,'' in {\em Proceedings of the IEEE
  International Conference on Quantum Computing and Engineering (QCE)}, (New
  York, NY, USA), pp.~292--300, 2021.

\bibitem{QAOA_ICQCE2}
A.~Kakkar, J.~Larson, A.~Galda, and R.~Shaydulin, ``Characterizing error
  mitigation by symmetry verification in qaoa,'' in {\em Proceedings of the
  IEEE International Conference on Quantum Computing and Engineering (QCE)},
  (Broomfield, CO, USA), pp.~635--645, 2022.

\bibitem{Csaba_Noise}
G.~Csaba and W.~Porod, ``Noise immunity of oscillatory computing devices,''
  {\em IEEE Journal on Exploratory Solid-State Computational Devices and
  Circuits}, vol.~6, no.~2, pp.~164--169, 2009.

\bibitem{SciPyProceedings_11}
A.~A. Hagberg, D.~A. Schult, and P.~J. Swart, ``Exploring network structure,
  dynamics, and function using networkx,'' in {\em Proceedings of the 7th
  Python in Science Conference}, (Pasadena, CA USA), pp.~11 -- 15, 2008.

\bibitem{gen_random_reg}
J.~H. Kim and V.~H. Vu, ``Generating random regular graphs,'' in {\em
  Proceedings of the Association for Computing Machinery (ACM)}, (New York, NY,
  USA), p.~213–222, 2003.

\bibitem{QAOA_CompleteGraph}
M.~Dupont, N.~Didier, M.~J. Hodson, J.~E. Moore, and M.~J. Reagor, ``The
  quantum approximate optimization algorithm performance with low entanglement
  and high circuit depth,'' {\em Phys. Rev. A}, vol.~106, p.~022423, 2022.

\bibitem{NelderMead}
J.~A. Nelder and R.~Mead, ``A simplex method for function minimization,'' {\em
  Comput. J.}, vol.~7, p.~308, 1965.

\bibitem{QAOA_NelderMead1}
L.~Zhou, S.-T. Wang, S.~Choi, H.~Pichler, and M.~D. Lukin, ``Quantum
  approximate optimization algorithm: Performance, mechanism, and
  implementation on near-term devices,'' {\em Phys. Rev. X}, vol.~10,
  p.~021067, 2020.

\bibitem{QAOA_NelderMead2}
X.~Lee, Y.~Saito, D.~Cai, and N.~Asai, ``Parameters fixing strategy for quantum
  approximate optimization algorithm,'' in {\em Proceedings of 2021 IEEE
  International Conference on Quantum Computing and Engineering (QCE)}, (New
  York, NY, USA), pp.~10--16, 2021.

\end{thebibliography}

\end{document}